\documentclass[12pt,a4paper]{article}
\usepackage[left=1in,right=1in]{geometry}
\usepackage{amsmath,amssymb}
\usepackage[pdfstartview=FitH,pdfpagemode=None]{hyperref}
\usepackage{cite}
\usepackage{braket}
\usepackage{xcolor}

\usepackage{authblk}
\renewcommand\Affilfont{\itshape\small}

\title{Semiclassical Quantization of the Superstring and Hagedorn Temperature}
            
\author[a]{Francesco Bigazzi\thanks{bigazzi@fi.infn.it}}
\author[a,b,c]{Tommaso Canneti\thanks{canneti@fi.infn.it}}            
\author[d,e]{Wolfgang M\"uck\thanks{mueck@na.infn.it}}

\affil[a]{Istituto Nazionale di Fisica Nucleare, Sezione di Firenze %
\protect\\ Via G. Sansone 1; 50019 Sesto Fiorentino (Firenze), Italy}
\affil[b]{Dipartimento di Fisica e Astronomia, Universit\`a di Firenze %
\protect\\ Via G. Sansone 1; 50019 Sesto Fiorentino (Firenze), Italy}
\affil[c]{Institute for Advanced Study, School of Natural Sciences%
\protect\\ Princeton, NJ 08540, USA}
\affil[d]{Dipartimento di Fisica ``Ettore Pancini'', Universit\`a degli Studi di Napoli ``Federico II''% 
\protect\\ Via Cintia; 80126 Napoli, Italy}
\affil[e]{Istituto Nazionale di Fisica Nucleare, Sezione di Napoli %
\protect\\ Via Cintia; 80126 Napoli, Italy}
\date{}

\numberwithin{equation}{section}

\def\be{\begin{equation}}
\def\ee{\end{equation}}
\def\bea{\begin{eqnarray}}
\def\eea{\end{eqnarray}}
\def\({\left (}
\def\){\right )}
\def\[{\left [}
\def\]{\right ]}

\newcommand{\ie}{i.e.,\ }
\newcommand{\eg}{e.g.,\ }

\newcommand{\rmd}{\,\mathrm{d}}

\newcommand{\e}[1]{\operatorname{e}^{#1}}

\newcommand{\flati}[1]{{\underline{#1}}}

\newtheorem{theorem}{Theorem}

\newcommand{\hide}[1]{}

\begin{document}
\maketitle

\begin{abstract}
{\small \noindent
In a recent paper \cite{Bigazzi:2022gal}, the semiclassical quantization of a string, winding once around the compact Euclidean time circle, 
on a supergravity background dual to the deep infrared regime of a confining finite temperature gauge theory, was carried out. %As a result, The main outcome of that work has been computing 
The string mass-shell condition and, by extrapolation, the Hagedorn temperature to leading order in the holographic limit was deduced. In this work, we improve on those results in three ways. First, we fix some missing details of the related light-cone quantization analysis. Second, we reconsider the problem under the lens of a background-covariant geometrical formalism. %which preserves the general covariance of the background. 
This allows us to put the semiclassical mass-shell condition on more solid grounds. Finally, %we after recalling that going to the Hagedorn regime requires going beyond the semiclassical limit, 
going beyond the semiclassical regime, we compute the Hagedorn temperature at next-to-leading order in the holographic limit. The sub-leading correction turns out to arise entirely from the contribution of the zero modes of the massive worldsheet scalar fields. Our result matches that of a recent analysis in the literature based on the Horowitz-Polchinski stringy star effective model. %The latter, in fact, is thought to be sensible in the near-Hagedorn limit, where the semiclassical string methods are not reliable. We show how the mass-shell condition for the string ground state can be recovered in the effective approach and we observe that the subleading correction to the Hagedorn temperature is due to the massive worldsheet scalar fields. 
%as due to the based on Despite the  to be valid infer methods allow us to reproduce the mass-shell condition in a filling some gaps which where missing  analysis carried Despite the outcome is correct, only the leading contribution in $\alpha'$ has to be considered reliable and details about the light-cone quantization have not been taken into accounts. In this work, we propose two versions of the rigorous computation fixing all the subtitles. The improvements make the first subleading contribution robust against one-loop corrections and deviations from the supergravity regime. This makes possible a comparison between our worldsheet method and other recent proposals in the literature.
}
\end{abstract}

\newpage
\tableofcontents

\linespread{1.3}

\section{Introduction}
\label{intro}

This work is motivated by the recent paper \cite{Bigazzi:2022gal} by two of us and A. L.~Cotrone, in which the semiclassical quantization of a closed string winding once around the compact Euclidean time direction and sitting in a nearly flat region of Witten's type-IIA supergravity background \cite{Witten:1998zw}, was considered. The region was chosen to holographically correspond to the deep infrared regime of the confining phase of the dual $SU(N)$ gauge theory, henceforth called Witten-Yang-Mills (WYM) theory. The holographic map holds in the $N\gg1$, $\lambda\gg1$ limits, $\lambda$ being the `t~Hooft coupling of the theory at a certain scale. Two central results of \cite{Bigazzi:2022gal} were the mass-shell condition for the string states and a prediction for the Hagedorn temperature of the WYM model, to leading order in $1/\sqrt{\lambda}$.   

The starting point behind \cite{Bigazzi:2022gal} and most of this paper is the statement, formulated in the seminal works \cite{Sathiapalan:1986db, Kogan:1987jd, Atick:1988si}, that the string Hagedorn temperature is the temperature at which the ground state of a closed string winding once around the thermal circle is massless. Beyond the Hagedorn temperature, a new tachyon would appear in the spectrum, signalling a phase transition. Extending this idea from the standard calculation in flat spacetime \cite{Atick:1988si} to strings on curved supergravity backgrounds is obviously extremely hard, foremost because quantization of the entire (super-)string is possible only in selected cases; Witten's background is certainly not among them. However a semiclassical approach appears feasible, and thus it is worth exploring it in detail. 
%from which, as long as higher order corrections are under control, one may expect correct predictions at least at the first few perturbative orders, where the meaning of ``few'' should be determined on a case-by-case basis. %In our case, it means leading and, to some extent, sub-leading. 

Unfortunately, due to subtleties hidden in the light-cone gauge quantization, some statements in \cite{Bigazzi:2022gal} are not accurate, although the above mentioned results, obtained with reasonable physical intuition, turn out to be correct.
In this paper, we improve on the calculation of \cite{Bigazzi:2022gal} in several respects. First, we fill the gap in the light-cone gauge computation including a careful treatment of the un-physical (gauge) modes and, in particular, of the light-cone gauge condition which removes them from the spectrum. These issues will be fixed in section~\ref{lightcone}. 
Second, in order to provide an independent confirmation of the above results and pave the way for other applications, we study the problem by means of a coordinate-invariant geometrical approach.\footnote{This approach must not be confused with the covariant quantization method in string theory, in which all spacetime coordinates are treated on equal footing. Rather, in this approach the string degrees of freedom are described by geometric objects in a coordinate-invariant manner.} This approach is based on the geometrical formalism pioneered in \cite{Alvarez-Gaume:1981exa}, which uses the exponential map and the geometry of embedded manifolds in order to parameterize the fluctuations around the classical background string worldsheet in a fully geometric and coordinate-independent manner. This method is well known in the AdS/CFT literature, see, \eg \cite{Faraggi:2011ge, Forini:2015mca}. The semiclassical quantization in the geometric approach will be presented in section~\ref{covariant}.  For the convenience of the reader and to explain the notation, a complete account of the method is included in appendices~\ref{embed} and \ref{exp}. 

Of course, the light-cone gauge and the geometrical approach lead to physically equivalent results, in particular to the same mass-shell condition. Both being semiclassical calculations at second order, they also share the same limitations.\footnote{In a semiclassical calculation, it is assumed that the quantum fluctuations are parametrically small with respect to the classical background parameters, without necessarily specifying the expansion parameter. In string theory, the expansion parameter is usually $\alpha'$; here it can be rephrased as $1/\sqrt{\lambda}$. The terms that are neglected are quartic or higher in the fluctuations, which means they are suppressed by $\mathcal{O}(1/\lambda)$ compared to the second-order terms.} A posteriori, the results turn out to be strictly valid only at relatively low temperatures.  %well below the Hagedorn temperature. However, it turns out that these results cannot be trusted for temperatures near the Hagedorn temperature. 
In fact, near the Hagedorn temperature some of the fluctuations are seen to grow to the same order of magnitude as the background parameters; or vice versa, some background parameters become parametrically small. This means that the perturbative expansion is not fully under control, signaling the failure of the semiclassical approach in that regime. Our third major improvement tries to address this issue. In section~\ref{TH}, we propose a new perturbative scheme for the string fluctuations, which takes into account the quantum correction to the classical mass-shell condition in the near-Hagedorn regime. This allows us to extract the mass-shell condition and deduce the Hagedorn temperature to next-to-leading order (NLO) in $1/\sqrt{\lambda}$. %Interestingly, the leading contribution coincides with the one already found, by extrapolation, in \cite{Bigazzi:2022gal}.
% (deleted, because the leading order result is a trivial result of the rescaling from the "stringy" coordinates to the WYM coordinates)
As we will show explicitly, the first sub-leading correction, which is entirely due to the zero modes of the massive worldsheet scalars, exactly matches with the sub-dominant contribution recently computed within an effective approach \cite{Urbach:2023npi} based on the Horowitz-Polchinski \cite{Horowitz:1997jc} string star model. 

To sum up, this work is organized as follows. Witten's background and the classical type IIA superstring solution, which are the starting points of the analysis carried out in sections~\ref{lightcone} and~\ref{covariant}, will be reviewed in section~\ref{background}. After presenting our results on the Hagedorn temperature in section~\ref{TH}, we will wrap up our results in the conclusion section~\ref{conc}. Appendices~\ref{embed}--\ref{app:T} contain review material and technical details of our computations.

\section{Background configuration and superstring action}
\label{background}

In this section, we provide details about the background dual to Witten's holographic Yang-Mills model %(WYM from now on)
and the reference classical string configuration. Although we try to maintain continuity with \cite{Bigazzi:2022gal}, we will adopt a slightly different notation in order to declutter the equations. A dictionary will be provided.

\subsection{Witten background}
\label{wbg:bulk.bg}

The type-IIA supergravity configuration dual to of the WYM theory, also known as the Witten background \cite{Witten:1998zw}, is given by\footnote{In the literature, there are several versions of the Witten background, differing from each other in the choice of coordinates. A popular form is used in \cite{Kruczenski:2003uq, Bigazzi:2004ze} with coordinates $x^\mu$ and the holographic coordinate $u\in[u_0,\infty)$ of dimension length. In \cite{Bigazzi:2022gal}, the $x^\mu$ have been made dimensionless via a rescaling by $m_0$. Here, we also introduce the dimensionless variable $v=u/u_0$.}
\begin{equation}
\label{wbg:bg}
\begin{aligned}
	\rmd s^2 &= m_0 R^3 \left[ v^{\frac32}  \left( \eta_{\mu\nu} \rmd x^\mu \rmd x^\nu + \frac49 f(v) \rmd \theta^2 \right) 
	+ \frac{\rmd v^2}{v^{\frac32} f(v)} + v^{\frac12} \rmd \Omega_4^2\right] ~, \\
	f(v)&= 1- v^{-3}~,\quad \e{\phi}= g_s (m_0R)^{\frac32} v^{\frac34}, \quad F_4 = 3 R^3 \omega_4~, \quad 
	R = (\pi g_s N)^{\frac13}\alpha'{}^{\frac12}  
\end{aligned}
\end{equation}
where $\eta=\text{diag}\{-,+,+,+\}$ and $\omega_4$ is the volume form of the unit $S^4$. The holographic coordinate $v\in[1,\infty)$ and $\theta\in[0,2\pi]$ parameterize a subspace with the shape of a cigar, the tip being located at $v=1$. Similar to polar coordinates, there is a coordinate singularity at $v=1$. In section~\ref{lightcone}, we shall introduce regular coordinates.
The parameter $m_0$ is the typical glueball mass scale in the WYM theory, $m_0 =\frac23 M_{\mathrm{KK}}$ in the notation of \cite{Kruczenski:2003uq}, where $M_{\mathrm{KK}}$ is the mass gap of the Kaluza-Klein tower for the compact $\theta$ direction. It relates the dimensionless coordinates $x^\mu$ to the physical, dimensionful coordinates of the dual WYM theory. The relation to the other gauge theory parameters is as follows \cite{Kruczenski:2003uq, Bigazzi:2004ze, Bigazzi:2022gal}.\footnote{Note that we adopt the convention of \cite{Bigazzi:2022gal} for $g_{YM}^2$, which differs by a factor of $2$ from the convention of \cite{Kruczenski:2003uq, Bigazzi:2004ze}.} The Yang-Mills coupling\footnote{In the holographic limit, with $\lambda\gg1$, the WYM model at low energy is a $3+1$ dimensional Yang-Mills theory coupled to the tower of Kaluza-Klein adjoint matter fields.} at the scale $M_{KK}$ is given by 
\begin{equation}
\label{wbg:g.YM}
	g_{YM}^2 = 4 \pi g_s \alpha'{}^{\frac12}\frac32 m_0~.  
\end{equation}
Thus, one can write the `t~Hooft coupling as
\begin{equation}
\label{wbg:thooft}
	\lambda = g_{YM}^2 N = 6 \pi g_s N \alpha'{}^{\frac12} m_0 =6 m_0 R (\pi g_s N)^{\frac23}~.  
\end{equation}
As observed in \cite{Kruczenski:2003uq} the supergravity solution is reliable, if 
$$g_{YM}^4 \ll \frac1{\lambda} \ll 1~,$$
which is certainly the case in the planar limit at large 't~Hooft coupling. 

In order to simulate a finite temperature, we will compactify a space-like direction, say $x^1$, on a circle as
\begin{equation}
\label{wbg:x1.compact}
	x^1 \sim x^1 + 2\pi \rho~.
\end{equation}
The dimensionless radius $\rho$ is related to the inverse physical temperature $\beta$ in the gauge theory by 
\begin{equation}
\label{wbg:rho.beta}
	2 \pi \rho = m_0 \beta~.
\end{equation}
Notice that \eqref{wbg:bg} has Lorentzian signature, which is essential for the Majorana condition on the spinors, with $x^0$ being the time direction.

\subsection{Background string at the tip of the cigar}
\label{bgstring}

A classical closed string winding once, with zero momentum, around the $x^1$ circle and sitting at the tip of the cigar is described by the following embedding:
\begin{equation}
\begin{aligned}
\label{wbg:ans}
	&x^1 = \rho \sigma~,\qquad &&x^i = p^i \tau  \qquad (i=0,2,3)~, \\
	&v =1 +\epsilon^2~, \qquad &&\text{$\theta$, $\Omega_4$ constant}~.   
\end{aligned}
\end{equation}
%
%where $e^i$ is a \emph{time-like} 3-d unit vector. This automatically satisfies the Virasoro constraint in the Polyakov formalism. 
The Virasoro constraint at the classical level requires
\be
M^2 =  \rho^2 \,,\quad M^2 \equiv -\eta_{ij}p^i p^j\,.
\label{classicalV}
\ee
As it will become clear in the following (see eq.~\eqref{massshellcond}), this condition receives quadratic corrections from the worldsheet field fluctuations around the classical configuration.
Moreover, we are free to adopt center-of-mass coordinates, $p^2=p^3=0$, which we shall do throughout section \ref{covariant}. 

Note that we have placed the string slightly off the tip of the cigar in order to avoid the coordinate singularity at $v=1$. One could avoid this by using regular coordinates as in section~\ref{lightcone}. However, once the geometrical quantities on the background worldsheet have been calculated, it is safe to set $\epsilon= 0$.
We will see in subsection~\ref{cov:dynamics} that the worldsheet of the embedding is minimal, \ie it is a classical solution, only for $\epsilon=0$.

\subsection{Type-IIA superstring in the Witten background} 

The type-IIA superstring in the Green-Schwarz formalism is described by the action,\footnote{For simplicity, we drop the couplings to those background fields, which are manifestly zero in the Witten background.} 
\begin{equation}
\label{str:action}
	S= S_P + S_F~,
\end{equation}
consisting of the Polyakov action
\begin{equation}
\label{str:Polyakov}
	S_P = -\frac1{4\pi\alpha'} \int \rmd^2 \sigma \sqrt{-h}\, h^{\alpha\beta} g_{\alpha\beta}~,
\end{equation}
and the fermion action (to quadratic order in the ten-dimensional Majorana spinor $\psi$) \cite{Martucci:2003gc}
\begin{equation}
\label{str:ferm.action}
	S_{F} = -i \int \rmd^2\sigma \sqrt{-h}\, \bar{\psi} \left( h^{\alpha \beta} + \epsilon^{\alpha\beta} \Gamma^{(11)} \right) \Gamma_\alpha D_\beta \psi~,
\end{equation}
where the Gamma matrices are such that $\Gamma^n = e^n_{. \, \flati{n}} \Gamma^\flati{n}$, $\left\{\Gamma^\flati{m}, \Gamma
^\flati{n} \right\} = 2 \eta^{mn} \mathbb{I}_{32}$. This implies that $\Gamma^\flati{0}$ is anti-hermitean and so $\bar\psi=\psi^{\dagger}\Gamma^{\underline 0}$. Here, $g_{\alpha\beta}$ is the induced metric on the string worldsheet,
\begin{equation}
\label{str:induced.metric}
	g_{\alpha\beta} = (\partial_{\alpha} x^m)(\partial_{\beta} x^n) g_{mn}~,
\end{equation}
while $h_{\alpha\beta}$ is an independent reference metric, which also determines the weight of the Levi-Civita tensor $\epsilon^{\alpha\beta}$.\footnote{For aesthetics, we prefer to work with tensors instead of densities. Of course, the weight $\sqrt{-h}$ cancels between the measure and $\epsilon^{\alpha\beta}$, so that this term is independent of $h_{\alpha\beta}$.} 
In the fermion action \eqref{str:ferm.action}, the covariant gamma matrices $\Gamma_\alpha$ contain the zwei-bein with respect to the induced metric $g_{\alpha\beta}$. Moreover, $\Gamma^{(11)}$ denotes the bulk chirality matrix, and $D_\alpha$ stands for the pull-back of the generalized covariant derivative \cite{Martucci:2003gc}
\begin{equation}
\label{str:Dn}
	D_n = \partial_n +\frac14 \omega_{n\flati{ab}} \Gamma^{\flati{ab}} - \frac1{8\cdot 4!} \frac{\e{\phi}}{g_s} F_{\flati{abcd}} \Gamma^{\flati{abcd}} \Gamma_n~. 
\end{equation}

Let us recall from \cite{Bigazzi:2022gal} that the worldsheet fermionic fields are taken to obey antiperiodic boundary conditions when $\sigma\rightarrow\sigma+2\pi$. Worldsheet bosons are instead periodic, as usual.

The action is invariant under worldsheet diffeomorphisms and under Weyl rescalings of $h_{\alpha\beta}$. These can be used to fix the conformal gauge
\begin{equation}
\label{str:conf.gauge}
	h_{\alpha\beta} = \eta_{\alpha\beta}~,
\end{equation}
after which the field equation for $h_{\alpha\beta}$ becomes the Virasoro constraint
\begin{equation}
\label{str:Virasoro}
  g_{\alpha\beta} - \frac12 h_{\alpha\beta} h^{\gamma\delta} g_{\gamma\delta} + 4\pi\alpha' i \bar{\psi} 
  \left( \Gamma_{(\alpha} D_{\beta)} - \frac12 h_{\alpha\beta} h^{\gamma\delta} \Gamma_\gamma D_\delta \right) \psi =0~.
\end{equation} 
\section{Light-cone gauge approach}
\label{lightcone}

Light-cone gauge quantization is a well-understood method of quantization for string theory in flat space and allows to work directly with the physical degrees of freedom in a non-covariant way. 

However, in curved space, the usual gauge-fixing has to be modified in a proper way in order to preserve the consistency of the theory. In particular, the stress-energy tensor crucially depends on the second order fluctuations around the classical configuration. As we will see, they are in turn fixed by the second order contributions to the equations of motion related to the total action of the theory. The latter represent an obstruction to the standard light-cone gauge choice and we have to deform it at second order. 

In this section, we will describe a fully consistent non-covariant semiclassical approach to the problem, solving the critical issues about the gauge-fixing which were omitted in \cite{Bigazzi:2022gal}.

\subsection{Worldsheet spectrum and gauge fixing}
\label{lightcone:spectrum}

Following a semiclassical approach, we consider small fluctuations around the classical string configuration \eqref{wbg:ans} and, to begin with, their linearized dynamics. The size of the fluctuations will turn out to be generically of order $1/\sqrt{\lambda}\ll1$; the latter is thus the effective expansion parameter within the semiclassical approach.\footnote{We do not explicitly expand in powers of $1/\sqrt{\lambda}$, but formally in powers of fluctuations around the classical background.} To avoid potential problems arising from the coordinate singularity at the tip of the cone, as anticipated in section \ref{bgstring}, we will be working in Cartesian coordinates $y_1, y_2$ replacing the ``polar'' coordinates $v,\theta$. The calculation can also be done in the polar coordinates; we include it in appendix~\ref{appUT} for completeness.  
Around the tip
\be \label{link}
v-1 = \frac34 (y_1^2 + y_2^2) \equiv \frac34 \, y^2\, 
\ee
and so the classical string is placed at the origin
\be \label{classicY}
y_1 = y_2 = 0 \, .
\ee
Small fluctuations around the classical configuration probe only the vicinity of the tip of the cigar. Working up to second order in the fluctuations, it is sufficient to keep the fluctuation terms in the metric components only for those directions in which the background string extends. All other components can be approximated by the corresponding tangent space metric. In particular, the $S^4$ sector is described by proper Cartesian coordinates $z^I$, with $I=6,7,8,9$. Classically we have $z^I=0$. Therefore, to quadratic order in $y\ll1$, the metric \eqref{wbg:bg} effectively reads 
\begin{align} \label{Wbacktipy}
	\rmd s^2 &\approx \frac{\alpha' \lambda}{6}\left\{ \[ 1 +\frac{9}{8} ( y_1^2 + y_2^2 ) \] \eta_{\mu\nu} \rmd x^\mu \rmd x^\nu +\rmd y_1^2 + \rmd y_2^2  + \rmd z^I \rmd z^I\right\}\\
\notag & \equiv \frac{\alpha' \lambda}{6} \tilde g_{mn}\rmd x^m \rmd x^{n} \, .
\end{align}
The prefactor is $m_0R^3=\frac16 \lambda \alpha'$. 

From the Polyakov action \eqref{str:Polyakov}, the linearized field equations are then found to be
\be
\label{eom.bosons}
\eta^{\alpha\beta}\partial_{\alpha}\partial_{\beta} x^{\mu}=0\,,\quad \left(\eta^{\alpha\beta}\partial_{\alpha}\partial_{\beta} - \mu^2\right)y_{i} = 0\,,\quad \eta^{\alpha\beta}\partial_{\alpha}\partial_{\beta}z^{I}=0\,,
\ee
where
\be
\mu^2 = \frac98 (M^2+\rho^2)\,,
\label{mugen}
\ee
is the dimensionless mass parameter of the modes $y_1, y_2$. Taking the classical Virasoro constraint \eqref{classicalV} into account, \eqref{mugen} becomes\footnote{The $\mathcal{O}(\lambda^{-1})$ corrections to the classical Virasoro constraint, which are due to the fluctuating worldsheet modes, would effectively produce, if inserted in \eqref{mugen}, subleading corrections to our quadratic worldsheet action.} 
\be
\label{mu.rho}
\mu = \frac32 \rho\,.
\ee
The linearized equations allow us to choose a ``first-order light-cone gauge", such that
\be
x^+ \equiv \frac{1}{\sqrt{2}}[x^0 + x^2] = p^+ \tau + \xi_{(2)}^+\,,
\label{flc}
\ee
where $\xi_{(2)}^+$ is a second-order correction quadratic in the field fluctuations, which, as it will become clear in the following, cannot be consistently set to zero. In this gauge, the physical bosonic degrees of freedom reduce to six transverse massless modes $x^1, x^3, z^I$ and two massive ones $y_1,y_2$.

The linearized equations for the fermionic modes reduce, after some algebra, to the condition
\be
\left[p^i \, \eta_{ i  j} \,\Gamma^{\underline j} + \rho \,\Gamma^{(11)}\Gamma^{\underline 0}\right]\left[\partial_{\tau}+\Gamma^{(11)}\partial_{\sigma}- \frac\mu2 \, \Gamma^{(11)} \, \Gamma^{\underline 1}\,\tilde\Gamma\right]\psi =0\,,
\label{fetot}
\ee
which implies
\be
\left[\partial_{\tau}+\Gamma^{(11)}\partial_{\sigma}- \frac\mu2 \,\Gamma^{(11)}\,\Gamma^{\underline{1}}\tilde\Gamma\right]\psi =0\,,
\label{fe}
\ee
where
\begin{equation}
\label{wbg:tilde.gamma}
	\tilde{\Gamma} = \Gamma^{\flati{6789}}
\end{equation}
is a Euclidean 4-d chirality matrix satisfying $\tilde{\Gamma}^2=1$. 
This in turn implies that
\be
\left(-\eta^{\alpha\beta}\partial_{\alpha}\partial_{\beta} + \frac{\mu^2}{4}\right)\psi=0\,.
\ee
As discussed in \cite{Bigazzi:2022gal, Bigazzi:2004ze}, these give rise to eight physical massive fermionic modes on the string worldsheet. See also the appendix \ref{appF} for further details on the fermionic sector of the model. 

\subsection{Virasoro constraint}
The equations of motion have to be supplemented with the Virasoro constraints, which amount to the vanishing of the stress-energy tensor
\be
{\mathcal T}_{\alpha\beta} = {\mathcal T}_{\alpha\beta}^B + {\mathcal T}_{\alpha\beta}^F\,,
\ee
where
\be
{\mathcal T}_{\alpha\beta}^B = \partial_{\alpha}x^m\partial_{\beta}x^n \tilde g_{mn} -\frac{1}{2}\eta_{\alpha\beta}\eta^{\gamma\delta}\partial_{\gamma}x^m\partial_{\delta}x^n \tilde g_{mn}\,,
\ee
and
\be
i \, {\mathcal T}_{\alpha\beta}^F = \frac12\bar\psi\Gamma_{\underline\mu}\partial_{\{\alpha}x^{\mu}\partial_{\beta\}}\psi - \frac{3}{8}\partial_{\alpha}x^{\mu}\partial_{\beta}x^{\nu} \eta_{\mu\nu} \bar\psi\widetilde\Gamma\psi -\frac{1}{2}\eta_{\alpha\beta}\eta^{\gamma\delta}\left[\bar\psi\Gamma_{\underline\mu}\partial_{\gamma}x^{\mu}\partial_{\delta}\psi - \frac{3}{8}\partial_{\gamma}x^{\mu}\partial_{\delta}x^{\nu}\eta_{\mu\nu}\bar\psi\widetilde\Gamma\psi\right]\,.
\ee
The stress-energy tensor, deduced as usual from the functional derivative of the action w.r.t.~the worldsheet metric\footnote{Here, we adopt a normalization such that ${\mathcal T}_{\alpha\beta} = - \frac{24\pi}{\lambda\sqrt{-h}} \, \delta S_P / \delta h^{\alpha\beta}$ and, at least in this section, we rescale the fermions as $\psi \mapsto i \sqrt{\lambda/(24\pi)} \psi$ in order to simplify the equations.}, is obviously quadratic in the field fluctuations. It is traceless and conserved
\be
\eta^{\alpha\beta}{\mathcal T}_{\alpha\beta}=0\,,\quad \partial_{\alpha}{\mathcal T}^{\alpha\beta}=0\,.
\ee
Stress tensor conservation is implied by the equations of motion for the bosonic and fermionic modes. Crucially, in order to work out this result, we need to include the second order corrections to the equations of motions of the modes $x^{\mu}$ which have non-trivial classical configurations.
These equations receive contributions from both the bosonic and the fermionic part of the action $S=S_P+S_F$. Using the classical solutions \eqref{wbg:ans} and the fermionic equations of motion (\ref{fe}), the second order equations read\footnote{These equations have to be satisfied order by order expanding $x^\mu$ around the classical configuration \eqref{wbg:ans} up to second order in quantum fluctuations. See next subsection for more details about it.}
\be \label{secordeomxmu}
%\eta^{\alpha\beta}\partial_{\alpha}\partial_{\beta} x^{\mu}= \frac{3}{2}m_0\alpha' p_i\widetilde\eta^{\mu i}\partial_{\tau}(y_iy_i) - \frac{3}{2}m_0\frac{\beta}{2\pi}\widetilde\eta^{\mu 0}\partial_{\sigma}(y_iy_i) +\frac{i}{2}\partial_{\alpha}F^{\mu\,\alpha}\,,
\partial_{\tau}\left[\partial_{\tau} x^{\mu} + \frac{9}{8} \, p_i \, \eta^{\mu i} \, y^2+ \frac{3}{8} i \, F^{\mu\,\tau}\right] = \partial_{\sigma}\left[\partial_{\sigma}x^{\mu}+\frac{9}{8} \, \rho \, \eta^{\mu 1} y^2 - \frac{3}{8} i \, F^{\mu\,\sigma}\right]\,,
\ee 
where we recall that $y^2\equiv y_1^2+y_2^2$ and
\begin{align}
&F^{\mu\,\tau}= \rho\,\bar\psi\,\Gamma^{(11)}\,\widetilde\Gamma\,\psi\,\eta^{\mu 1} + p_i \, \eta^{\mu i}\,\bar\psi\,\widetilde\Gamma\,\psi\,,\nonumber \\
&F^{\mu\,\sigma}= \rho \, \bar\psi\left[\Gamma^{\underline\mu}\Gamma^{\underline1}-\eta^{\mu1}\right]\widetilde\Gamma\psi+ p^i\,\eta_{ij}\,\bar\psi\,\Gamma^{(11)}\,\Gamma^{\underline\mu \underline j}\,\widetilde\Gamma\,\psi\,.
\end{align}
Notice that, being quadratic in the fermionic fields, $F^{\mu\,\tau}$ and $F^{\mu\,\sigma}$ are periodic in $\sigma$.

From the equations above we deduce, in particular, that the light-cone gauge choice $x^+=p^+\tau$ cannot be implemented to second order. 
However, the ``first-order light-cone gauge'' introduced in (\ref{flc}) is enough to express the (quadratic order) string mass-shell and level matching conditions in terms of the number operators of the eight transverse bosonic and fermionic modes.

\subsection{Mass-shell condition}
The string mass-shell condition emerges from the integral relation
\be
\int_{0}^{2\pi} \rmd\sigma\, \delta^{\alpha\beta}\, {\mathcal T}_{\alpha\beta} =0\,.
\ee
Let us see how it looks like in detail, expanding the stress-energy tensor to quadratic order in the field fluctuations. Let us write
\be \label{secondorderex}
x^{\mu} = X^{\mu} + \xi_{(1)}^{\mu} + \xi_{(2)}^{\mu}\,, \qquad (\xi_{(1)}^+=0)\,,
\ee
where $X^{\mu} = \rho \sigma \delta^{\mu1} + \sum_i p^i \tau \delta^{\mu i} $ are the classical solutions given in \eqref{wbg:ans} and $\xi_{(1)}^{\mu}, \xi_{(2)}^{\mu}$ are first and second order fluctuations. For what concerns the other bosonic fields, which are zero at the classical level, we will keep calling $z^I, y_i$ the related first order fluctuations. 

According to the results collected above we have
\bea
&\eta^{\alpha\beta}\partial_{\alpha}\partial_{\beta}\xi_{(1)}^{\mu}=0\,,\nonumber \\
&\partial_{\tau}\left[\partial_{\tau} \xi_{(2)}^{\mu} + \frac{9}{8} \, p_i\, \eta^{\mu i} y^2+ \frac{3}{8} i \, F^{\mu\,\tau}\right] = \partial_{\sigma}\left[\partial_{\sigma}\xi_{(2)}^{\mu}+\frac{9}{8} \, \rho \, \eta^{\mu 1} y^2 - \frac{3}{8} i \, F^{\mu\,\sigma}\right]\,.
\eea
Let us notice, in particular, that the latter equation and the periodicity condition of the closed superstring, imply that
\be
\partial_{\tau}\int_{0}^{2\pi}\rmd\sigma \left[ \partial_{\tau} \xi_{(2)}^{\mu} + \frac{9}{8} \, p_i\, \eta^{\mu i} y^2+ \frac{3}{8} i \, F^{\mu\,\tau} \right] =0\,
\ee
so that
\be
\int_{0}^{2\pi}\rmd\sigma \left[ \partial_{\tau} \xi_{(2)}^{\mu} + \frac{9}{8} \, p_i\, \eta^{\mu i} y^2+ \frac{3}{8} i \, F^{\mu\,\tau} \right] = 0\,,
\label{inter}
\ee
where we have put to zero the constant on the r.h.s. requiring that our system has the correct flat spacetime limit. 

Focusing for a moment on the bosonic part of the stress-energy tensor we get, in the first-order light-cone gauge (\ref{flc}) that, to quadratic order
\be
\delta^{\alpha\beta}{\mathcal T}^B_{\alpha\beta} = -M^2 + \rho^2 + 2\delta^{\alpha\beta}\partial_{\alpha} X^{\mu}\partial_{\beta}\left(\xi_{(1)}^{\nu} + \xi_{(2)}^{\nu}\right)\eta_{\mu\nu}+\delta^{\alpha\beta} {\mathcal T}_{\alpha\beta}^{t}\,,
\label{expa}
\ee
where ${\mathcal T}_{ab}^{t}$ accounts for the contribution of  the first order fluctuations of the eight transverse fields 
\be
\delta^{\alpha\beta} {\mathcal T}_{\alpha\beta}^{t}\equiv \delta^{\alpha\beta}\partial_{\alpha}\xi_{(1)}^p\partial_{\beta}\xi_{(1)}^q\delta_{pq}\,,
\label{resw}
\ee
with $\xi_{(1)}^p= \left(\xi_{(1)}^1, \xi_{(1)}^3, y_i, z_I \right)$.

Using the results collected above, and in particular eq. (\ref{inter}) we thus get\footnote{Notice that only the term involving $\partial_{\beta}\xi^{\nu}_{(2)}$ in (\ref{expa}) contributes to (\ref{boseT}): $\int \rmd\sigma \delta^{\alpha\beta}\partial_{\alpha} X^{\mu}\partial_{\beta}\xi_{(1)}^{\nu}\widetilde\eta_{\mu \nu}$ vanishes, since $\xi^{+}_{(1)}=0$ and the other $\xi^{\mu}_{(1)}$ fields have the same mode expansions as in flat space.}
\be
\int_{0}^{2\pi} \rmd\sigma\,\delta^{\alpha\beta}{\mathcal T}^B_{\alpha\beta} = 2\pi \left[-M^2 + \rho^2\right] + \int_{0}^{2\pi} \rmd\sigma \left[ \delta^{\alpha\beta} {\mathcal T}_{\alpha\beta}^{t}+ \mu^2 y^2 + \frac{i}{3}\mu^2\bar\psi\widetilde\Gamma\psi\right]\,.
\label{boseT}
\ee
%Remarkably, the combination $\delta_{\alpha\beta} {\mathcal T}_{\alpha\beta}^{t}+ \mu^2 y^2$ is proportional to the canonical Hamiltonian density for the transverse bosonic modes. 
Considering the fermionic contribution, we get, on shell,
\be
\int_{0}^{2\pi} \rmd\sigma\,\delta^{\alpha\beta}{\mathcal T}^F_{\alpha\beta} = \int_{0}^{2\pi} \rmd\sigma \left[-i \, p^i \, \eta_{ij} \, \bar\psi \,  \Gamma^{\underline j} \, \partial_{\tau}\psi -i \, \rho \, \bar\psi\, \Gamma^{(11)}\Gamma^{\underline 1} \, \partial_{\tau}\psi- \frac{i}{3}\mu^2\bar\psi \,\widetilde\Gamma \, \psi\right]\,.
\ee
Notice that the last term under integration, precisely cancels with the corresponding one in \eqref{boseT}. 

All in all we get the mass-shell condition
\be
\label{mass.shell}
0=\int_{0}^{2\pi} \rmd\sigma\,\delta^{\alpha\beta}{\mathcal T}_{\alpha\beta} =  2\pi \left[-M^2 + \rho^2 \right] + \frac{24\pi}{\lambda}\int_{0}^{2\pi} \rmd\sigma \left[{\cal H}^B + {\cal H}^F\right]\,,
\ee
where
\be
{\cal H}^B = \frac{\lambda}{24\pi} \left[\delta^{\alpha\beta} {\mathcal T}_{\alpha\beta}^{t}+ \mu^2 y^2\right]
\label{HamB}
\ee
and
\be
{\cal H}^F=  -i\frac{\lambda}{24\pi}\,\bar\psi\left[p^i \eta_{ij}\Gamma^{\underline j} + \rho \, \Gamma^{(11)} \Gamma^{\underline{1}}\right]\partial_{\tau}\psi
\label{hamf}
\ee
are, respectively, the canonical Hamiltonian densities for the bosonic and fermionic modes. After writing the Hamiltonian in normal-ordered form, \eqref{mass.shell} becomes 
\be
\label{massshellcond}
M^2= \rho^2 +\frac{12}{\lambda}\left[ N +\widetilde N - \Delta(\mu)\right] \,,
%  m_0^2 \frac{\beta^2}{4\pi^2}
\ee
where $N$ and $\widetilde N$ denote the level operators whose explicit expressions can be found in \cite{Bigazzi:2022gal},\footnote{They are also contained in the oscillator terms in \eqref{appT:L0.full} and \eqref{appT:tildeL0.full}.} and the function $\Delta(\mu)$ collects normal ordering contributions from the scalars and spinors,
\begin{equation}
\label{Delta:D.def}
	\Delta(\mu) = -6 \sum\limits_{n=1}^\infty n - \mu - 2 \sum\limits_{n=1}^\infty \sqrt{n^2+\mu^2} + 8 \sum\limits_{n=1}^\infty \sqrt{\left(n-\frac12\right)^2+\frac{\mu^2}{4}}~.
\end{equation}
Although the single sums in \eqref{Delta:D.def} are divergent, a standard calculation using Zeta-function regularization (see appendix~\ref{Delta}) leads to a finite expression for $\Delta(\mu)$. The mass matching condition \eqref{wbg:susy.mass.formula} is crucial in this context. We emphasize that the term $\sim\mu$ is the contribution of the zero modes of the two massive scalar fields $y_1,y_2$. 

The mass-shell condition \eqref{massshellcond} makes it clear how the classical Virasoro constraint $M^2=\rho^2$, receives ${\cal O}(\lambda^{-1})$ quantum corrections from the quadratic terms in the worldsheet fields. As we have already stressed, the latter are in fact ${\cal O}(\lambda^{-1/2})$ fluctuations around the classical background. We defer a further discussion of the result to section \ref{TH}.

\section{Superstring fluctuations -- geometric approach}
\label{covariant}

Our aim in this section is to find and quantize the fluctuations of the superstring degrees of freedom around the classical background worldsheet described in subsection~\ref{bgstring}. Here we use a coordinate-invariant geometrical formalism, which builds on the geometry of embedded manifolds and the exponential map reviewed in appendices~\ref{embed} and \ref{exp}, respectively. For our purposes, it is sufficient to work up to second order in the fluctuations.
The presentation will be independent from the analysis in the preceding section, but we will see that the final result for the mass-shell condition will be the same. 

An outline of this section is as follows. In the geometric method, the independent degrees of freedom are given by eight real scalars, $\chi^i$ ($i=2,3,\ldots,9$), parameterizing the fluctuations orthogonal to the background worldsheet and eight 2-d Majorana spinors, $\psi^s$ ($s=1,2,\ldots,8$). The tangential fluctuations are parameterized by a 2-d vector, $\zeta^\alpha$, which is subject to the Virasoro constraint and embodies some gauge freedom. In subsection~\ref{cov:dynamics}, we present the actions governing the dynamics of the scalars and the fermions. The field equations are solved and explicit mode expansions for the scalars and fermions will be given in subsections~\ref{cov:scalars} and \ref{cov:ferm}, respectively. In subsection~\ref{cov:vira}, we show how the Virasoro constraint determines the 2-d vector. Then, as an intermezzo, we present in subsection~\ref{cov:hagedorn.flat} the calculation of the mass-shell condition in flat spacetime, in which the expansion in fluctuations up to second order is exact. The result obviously agrees with the known expression. %, both for the type-II superstring and for the bosonic string. 
Finally, the mass-shell condition in WYM theory is calculated in subsection~\ref{cov:hagedorn}.

\subsection{Superstring fluctuations}
\label{cov:dynamics}

In this subsection, we shall write down the actions that govern the dynamics of the independent fields. For the scalars, the procedure for the expansion to second order is described in appendix~\ref{exp:string}, to which we refer for details. The resulting action is 
\begin{equation}
\label{str:S.P.expand}
	S_{P} = -\frac1{4\pi\alpha'} \int \rmd^2 \sigma \sqrt{-g} \left[ (\nabla_\alpha \chi_i)(\nabla^\alpha \chi^i) 
	 + \left( \mathcal{M}_{ij} - H_{i\alpha\beta} H_j{}^{\alpha\beta} \right) \chi^i \chi^j \right]~,
\end{equation} 
where $H_{i\alpha\beta}$ are the second fundamental forms of the background embedding, and
\begin{equation}
\label{str:mass.mat}
	\mathcal{M}_{ij} = -R_{mpnq} x^{\alpha m} x_{\alpha}^n N_i^p N_j^q~.
\end{equation}
We remark that \eqref{str:S.P.expand} has been obtained using the Virasoro constraint at \emph{linear} order to manipulate the quadratic terms in the action and dropping total derivative terms. In this way, the vector $\zeta^\alpha$ explicitly disappears from the action, but it remains as a constrained field, which we will deal with in subsection~\ref{cov:vira}.

In the action~\eqref{str:S.P.expand}, there are several ways in which the scalars couple to the background geometry. First, the covariant derivatives $\nabla_\alpha$ contain, besides the worldsheet connections (minimal coupling), connections in the normal bundle, see \eqref{embed:weingarten}. Second, the ``mass'' matrix arises from the spacetime curvature and extrinsic curvature components of the worldsheet.\footnote{In our case, the ``mass'' matrix is diagonal and constant, but in general it is not.} 
Therefore, to make \eqref{str:S.P.expand} explicit, it is necessary to calculate the geometric quantities characterizing the embedding \eqref{wbg:ans} of the background worldsheet, see appendix~\ref{embed}. Recall from subsection~\ref{bgstring} that, using polar coordinates around the tip of the cone, we consider a worldsheet at constant $v=1+\epsilon^2$ and must set $\epsilon\to 0$ only at the end to avoid the coordinate singularity. The tangent vectors directly follow from \eqref{wbg:ans}, taking into account the choice of center-of-mass coordinates and the conformal gauge ($p^0=\rho$, $p^2=p^3=0$),
\begin{equation}
\label{wbg.tang}
	x_\tau = \rho \frac{\partial}{\partial x_0}~, \quad x_\sigma = \rho \frac{\partial}{\partial x_1}~.
\end{equation}
The induced metric is
\begin{equation}
\label{wbg:g.induced}
	g_{\alpha\beta} = m_0 R^3 v^{\frac32} \rho^2 \eta_{\alpha\beta}~.
\end{equation}
To proceed, we need to choose a basis of eight normal vectors, $N_i^m$. Let us label them with the integers $i=2,3,\ldots, 9$ and take\footnote{In this notation, we reserve the indices $0$ and $1$ for the worldsheet directions. The coefficients in $N_4$ and $N_5$ are an artifact of the coordinate singularity at $v=1$.}
\begin{equation}
\label{wbg:normals}
\begin{aligned}
	N_{2,3} &= \frac{1}{(m_0R^3)^\frac12v^{\frac34}} \frac{\partial}{\partial x_{2,3}}~, &
	N_4 &= \frac{3v^{\frac34}}{2(m_0R^3)^\frac12\sqrt{v^3-1}} \frac{\partial}{\partial \theta}~, \\
	N_5 &= \frac{\sqrt{v^3-1}}{(m_0R^3)^\frac12v^{\frac34}}\frac{\partial}{\partial v}~,\qquad &
	N_{6,7,8,9} &= \frac{1}{(m_0R^3)^\frac12v^{\frac14}} e_{6,7,8,9}~,
\end{aligned}
\end{equation}
with $e_{6,7,8,9}$ a complete set of unit vectors on the unit $S^4$.

With these vectors at hand, one calculates the second fundamental forms (see appendix~\ref{embed:structure}). It turns out that the only non-zero components for general $v$ are
\begin{equation}
\label{wbg:H5}
	H_{5\tau}{}^\tau = H_{5\sigma}{}^\sigma = - \frac{3 \sqrt{v^3-1}}{4(m_0R^3)^\frac12v^{\frac74}}~, \qquad
	H_{5\alpha}{}^\alpha = - \frac{3 \sqrt{v^3-1}}{2(m_0R^3)^\frac12v^{\frac74}}~.
\end{equation}
In the second equation, we have explicitly written the trace in order to show that it vanishes only for $v=1$. Therefore, the classical string, whose field equation is \eqref{exp:H.field.equation}, $H_{i\alpha}{}^\alpha=0$, is bound to the tip of the cigar. Moreover, also the single components vanish there, so that we have, after letting $\epsilon\to0$,
\begin{equation}
\label{wbg:H.is.0}
	H_{i\alpha}{}^\beta = 0~.
\end{equation}
The connections in the normal bundle vanish, 
\begin{equation}
\label{wbg:A.is.0}
	A_{ij\alpha} = 0~.
\end{equation}

The fermion action is simply obtained from \eqref{str:ferm.action} by replacing $h_{\alpha\beta}$ by the background $g_{\alpha\beta}$ (using the Virasoro constraint for the background),\footnote{Now, $\epsilon^{\alpha\beta}$ includes the weight with respect to $g_{\alpha\beta}$.} 
\begin{equation}
\label{str:ferm.action.g}
	S_{F} = -i \int \rmd^2\sigma \sqrt{-g}\, \bar{\psi} \left( g^{\alpha \beta} + \epsilon^{\alpha\beta} \Gamma^{(11)} \right) \Gamma_\alpha D_\beta \psi~.
\end{equation} 
This can be re-written, after gauge-fixing the $\kappa$-symmetry, in terms of eight 2-d Majorana spinors. We postpone this reduction to subsection~\ref{cov:ferm}.

\subsection{Scalars}
\label{cov:scalars}

Taking into account \eqref{wbg:H.is.0}, the action \eqref{str:S.P.expand} gives rise to the field equations
\begin{equation}
\label{wbg:eom.scal}
	\nabla^\alpha\nabla_\alpha \chi^i  - \mathcal{M}^i{}_j\chi^j =0~,
\end{equation}
with the ``mass'' matrix $\mathcal{M}^i{}_j$ given by \eqref{str:mass.mat}. Recall the background induced metric \eqref{wbg:g.induced} with $v=1$ and \eqref{wbg:A.is.0}, which imply $\nabla_\alpha=\partial_\alpha$. For $\mathcal{M}^i{}_j$  one finds only two non-zero components,
\begin{equation}
\label{wbg:M55}
	\mathcal{M}^4{}_4 = \mathcal{M}^5{}_5 = \frac9{4m_0R^3}~.
\end{equation}
Putting everything together, \eqref{wbg:eom.scal} gives rise to six massless scalars and two massive ones,
\begin{align}
\label{wbg:eom.massivless}
	\left( -\partial_\tau^2 +\partial_\sigma^2 \right) \chi^{2,3,6,7,8,9} &=0~,\\
\label{wbg:eom.massive}
	\left( -\partial_\tau^2 +\partial_\sigma^2 - \mu^2 \right) \chi^{4,5} &=0~,
\end{align}
where the dimensionless mass parameter is $\mu=\frac32 \rho$ as in \eqref{mu.rho}. 
%is  given in
%
%\begin{equation}
%\label{wbg:mu.def}
%	\mu = \frac32 \rho~.
%\end{equation}

The mode expansion and quantization of the scalars is a standard procedure. For the massless scalars ($i=2,3,6,7,8,9$), we have
\begin{equation}
\label{wbg:scal.mode.massless}
	\chi^i = \chi^i_0 + \sqrt{2\alpha'} q^i_0 \tau + i \sqrt{\frac{\alpha'}{2}} \sum\limits_{n\neq 0} \left[ \frac{\alpha^i_n}{n} \e{-in(\tau-\sigma)}
	+ \frac{\tilde{\alpha}^i_n}{n} \e{-in(\tau+\sigma)} \right]~,
\end{equation}
with the commutators 
\begin{equation}
\label{wbg:scal.comm.massless}
	[\chi^i_0, q_0^j ] = i \delta^{ij}~, 
	\qquad [\alpha_m^i, \alpha_n^j] = [\tilde{\alpha}_m^i, \tilde{\alpha}_n^j] = n \delta^{ij} \delta_{m,-n}~,
\end{equation}
the reality condition $\alpha_{-n}^i = (\alpha_n^i)^\dagger$ and normal ordering defined as $\alpha_n^\dagger$ to the left of $\alpha_n$ for $n>0$. 

For the massive scalars ($i=4,5$), we use the somewhat simpler expansion
\begin{equation}
\label{wbg:scal.mode.massive}
	\chi^i = i \sqrt{\frac{\alpha'}{2}} \sum\limits_{n=-\infty}^\infty \frac1{\sqrt{\omega_n}} \left[ a_n^i \e{-i(\omega_n\tau+n\sigma)}
	- a_n^i{}^{\dagger} \e{i(\omega_n\tau+n\sigma)}\right]
\end{equation}
with 
\begin{equation}
\label{wbg:omega.n}
	\omega_n = \sqrt{n^2 +\mu^2}
\end{equation}
and the standard commutators
\begin{equation}
\label{wbg:scal.comm.massive}
	[a_m^i, a_n^j{}^\dagger] = \delta^{ij} \delta_{m,n}~.
\end{equation}

\subsection{Fermions}
\label{cov:ferm}

Consider the fermion action \eqref{str:ferm.action.g}. To write $D_\alpha$ explicitly, we use \eqref{embed:D} and note again that the second fundamental forms, the connections in the normal bundle, and the spin connections vanish on the background worldsheet. Moreover, to write out the term with the four-form in $D_\alpha$, we need to express $F_4$ in terms of the volume form of the physical $S^4$ in the background \eqref{wbg:bg} (not the unit $S^4$), $F_{[6789]} = \frac{3}{m_0^2R^3v} \epsilon_{6789}$. This gives
\begin{equation}
\label{wbg:Dalpha}
	D_\alpha = \partial_\alpha - \frac3{8 (m_0R^3)^\frac12 v^{\frac14}} \tilde{\Gamma} \Gamma_\alpha~,
\end{equation}
where $\tilde{\Gamma} = \Gamma^{\flati{6789}}$, as in \eqref{wbg:tilde.gamma}. Because the worldsheet does not extend along the $S^4$, $\Gamma^\alpha$ and $\tilde{\Gamma}$ commute. We also have $\epsilon^{\alpha\beta} \Gamma_\alpha = - \Gamma^{\flati{01}} \Gamma^\beta$ in two dimensions. Fixing the $\kappa$-symmetry is trivial, because the projector in \eqref{str:ferm.action.g} already removes half of the 32 spinor components with the condition
\begin{equation}
\label{wbg:kapp.fix}
	\Gamma^{\flati{01}}  \psi =  \Gamma^{(11)} \psi~.
\end{equation} 
This leaves 16 spinor components, which combine into eight 2-d Majorana spinors $\psi^s$, $s=1,2,\ldots, 8$. They can be chosen to be eigen-spinors of $\tilde{\Gamma}$, with eigenvalues $(-1)^s$.

Putting all together, the action \eqref{str:ferm.action.g} becomes\footnote{The factor in front is irrelevant for the fermion field equations, but matters for the normalization of the quantized field. It arises from converting $\Gamma_\alpha$ into $\Gamma_\flati{\alpha}$ and a factor of $2$ from the two terms in the projector. Note that the spinors are dimensionful.}  
\begin{equation}
\label{wbg:ferm.action.2}
	S_{F} = -i\,2\rho (m_0R^3)^\frac12 \int \rmd^2 \sigma\, \bar{\psi}^s \left( \Gamma^\flati{\alpha} \partial_\alpha - \mu_s \right) \psi^s~,
\end{equation}
where the sum over $s$ is implicit and the (dimensionless) masses are  
\begin{equation}
\label{wbg:mu.s}
	\mu_s = \frac12 (-1)^s \mu
\end{equation}
with $\mu$ given by \eqref{mu.rho}. As a nice check, one can verify that, as already observed in \cite{Bigazzi:2022gal},
\begin{equation}
\label{wbg:susy.mass.formula}
\sum_{b} m_b^2 = 2 \mu^2 = \sum_{f} m_f^2~,
\end{equation}
where the sums are over all bosonic and fermionic fields, respectively.

To proceed, let us choose the gamma matrices as 
\begin{equation}
\label{wbg:gamma.mat}
	\Gamma^{\flati{\tau}} = 
		\begin{pmatrix}
		0 &1 \\ -1 &0
		\end{pmatrix}~, \qquad
	\Gamma^{\flati{\sigma}} = 
		\begin{pmatrix}
		0 &1 \\ 1 &0
		\end{pmatrix}~.
\end{equation}
Consider a spinor $\psi^s$ that satisfies the Dirac equation $\left( \Gamma^\flati{\alpha} \partial_\alpha - \mu_s \right) \psi^s=0$ with anti-periodic boundary conditions along the circle. Its mode expansion is
\begin{equation}
\label{wbg:spin.mode}
	\psi^s = \sum\limits_{r\in \mathbb{Z}+\frac12} A_r \left[ \begin{pmatrix} -i(\omega_r+r)\\ \mu_s \end{pmatrix} c^s_r \e{-i(\omega_r \tau + r \sigma)} + \begin{pmatrix} \mu_s \\ i(\omega_r+r) \end{pmatrix} d^s_r \e{-i(\omega_r \tau - r \sigma)} \right]~,
\end{equation}
with 
\begin{equation}
\label{wbg:omega.r} 
	\omega_r = \operatorname{sgn} r \sqrt{r^2 +\frac14 \mu^2}~,
\end{equation}
the reality conditions $A_{-r}=A_r^\ast$, $c^s_{-r} =c^s_r{}^\dagger$, $d^s_{-r} =d^s_r{}^\dagger$, the standard anti-commutation relations 
\begin{equation}
\label{wbg:spin.acomm}
	\left\{ c^s_r, c^{s'}_{r'}\right\} = \left\{ d^s_r, d^{s'}_{r'}\right\} = \delta^{ss'} \delta_{r,-r'}~,  
\end{equation}
normal ordering defined by writing creation operators $c_r^s{}^\dagger$ and $d_r^s{}^\dagger$ to the left of the annihilation operators $c_r^s$ and $d_r^s$ for $r>0$,\footnote{To determine which of $c_r$ or $c_{-r}$ is to be interpreted as the annihilation operator, one considers the canonical  Hamiltonian and demands that each field quantum contributes an energy $|\omega_r|$. This also determines the normalization constant $A_r$.} and the normalization constants $A_r$ satisfying 
\begin{equation}
\label{wbg_spin.Ar}
	|A_r|^2 = \frac{1}{16\pi\rho (m_0R^3)^\frac12 \omega_r(\omega_r+r)}~.
\end{equation}

\subsection{Virasoro constraint}
\label{cov:vira}

In the two previous subsections, we have dealt with the dynamics of the unconstrained fields, \ie the scalars parameterizing the normal fluctuations and the spinors. The remaining vector $\zeta^\alpha$, which describes the tangential fluctuations along the worldsheet, is subject to the Virasoro constraint \eqref{str:Virasoro} and has otherwise no independent field equation. To second order in the fluctuations and dropping the terms that evidently vanish by the background relations, in particular \eqref{wbg:H.is.0}, the Virasoro constraint reads
\begin{align}
\label{wbg:Virasoro}
	2\nabla_{(\alpha}\zeta_{\beta)} - g_{\alpha\beta} \nabla_\gamma \zeta^\gamma 
	+ (\nabla_\alpha \zeta_\gamma)(\nabla_\beta \zeta^\gamma) - \frac12 g_{\alpha\beta} (\nabla_\gamma \zeta_\delta)(\nabla^\gamma \zeta^\delta) 
	+T_{\alpha\beta} =0~,
\end{align}
where $T_{\alpha\beta}= T^{(B)}_{\alpha\beta}+ T^{(F)}_{\alpha\beta}$ contains the contributions from the independent fields,
\begin{equation}
\label{wbg:TB}
	T^{(B)}_{\alpha\beta} = (\nabla_\alpha \chi^i)(\nabla_\beta \chi_i) - \frac12 g_{\alpha\beta} (\nabla_\gamma \chi_i)(\nabla^\gamma \chi^i) 
	- R_{mpnq}N_i^p N_j^q \chi^i \chi^j \left( x_\alpha^m x_\beta^n - \frac12 g_{\alpha\beta} x_\gamma^m x^{\gamma n} \right)
\end{equation}
and 
\begin{equation}
\label{wbg:TF}
	T^{(F)}_{\alpha\beta} = 4 \pi \alpha' i \bar{\psi}^s \Gamma_{(\alpha} D_{\beta)} \psi^s~.
\end{equation}

It is a crucial observation that \eqref{wbg:TB} and \eqref{wbg:TF} do not coincide, as far as the massive fields are concerned, with the canonical stress-energy tensors for free scalars and fermions, respectively. In fact, \eqref{wbg:TB} and \eqref{wbg:TF} are traceless as a consequence of the Weyl invariance of the original action,\footnote{This is evident for \eqref{wbg:TB}, while it follows from the Dirac equation in the case of \eqref{wbg:TF}.} while they are not conserved. The opposite is true for the canonical stress-energy tensors of massive fields. This means, in particular, that the divergence of the Virasoro constraint \eqref{wbg:Virasoro} cannot be shown to vanish by means of the field equations, also because there are no field equations for $\zeta^\alpha$ independently from the constraint. This fact somewhat complicates the procedure with respect to the case of flat spacetime, in which the physical excitations are massless, so that $T_{\alpha\beta}$ is both conserved and traceless. Therefore, in the case of flat spacetime, taking the divergence of \eqref{wbg:Virasoro} shows that $\zeta^\alpha$ must be harmonic, but this is not the case in general. To isolate an harmonic part we must proceed as follows. 

First, let us introduce a conserved tensor $\tilde{T}_{\alpha\beta}$ such that
\begin{equation}
\label{wbg:T.tilde}
	T_{\alpha\beta} = \tilde{T}_{\alpha\beta} - \frac12 g_{\alpha\beta} \tilde{T}~,  \qquad \nabla^\alpha\tilde{T}_{\alpha\beta}=0~.
\end{equation}
Specifically, we have $\tilde{T}_{\alpha\beta}=\tilde{T}^{(B)}_{\alpha\beta}+\tilde{T}^{(F)}_{\alpha\beta}$, where $\tilde{T}^{(B)}_{\alpha\beta}$ and $\tilde{T}^{(F)}_{\alpha\beta}$ are the canonical stress-energy tensors for the fluctuation fields $\chi$ and $\psi$, respectively. These are, of course, not traceless for the massive fields. 

Second, let us decompose $\zeta^\alpha$ into
\begin{equation}
\label{wbg:zeta.decomp}
	\zeta^\alpha = \zeta^{(1)\alpha} + \zeta^{(2)\alpha}~,  
\end{equation}
where $\zeta^{(1)\alpha}$ and $\zeta^{(2)\alpha}$ stand for the terms of first and second order in $\alpha'{}^\frac12$, respectively. 
Thus, \eqref{wbg:Virasoro} decomposes into
\begin{align}
\notag
	2\nabla_{(\alpha}\zeta^{(1)}_{\beta)} - g_{\alpha\beta} \nabla_\gamma \zeta^{(1)\gamma} 
	+2\nabla_{(\alpha}\zeta^{(2)}_{\beta)} - g_{\alpha\beta} \nabla_\gamma \zeta^{(2)\gamma} &\\
\label{wbg:Virasoro2}
	+ (\nabla_\alpha \zeta^{(1)}_\gamma)(\nabla_\beta \zeta^{(1)\gamma}) - \frac12 g_{\alpha\beta} (\nabla_\gamma \zeta^{(1)}_\delta)(\nabla^\gamma \zeta^{(1)\delta}) 
	&=- T_{\alpha\beta}~.
\end{align}
Because $T_{\alpha\beta}$ is of second order, $\zeta^{(1)\alpha}$ is a conformal Killing vector satisfying
\begin{equation}
\label{wbg:conf.Killing}
	2\nabla_{(\alpha}\zeta^{(1)}_{\beta)} - g_{\alpha\beta} \nabla_\gamma \zeta^{(1)\gamma} =0~,
\end{equation}
which implies $\Box \zeta^{(1)\alpha}=0$. Its existence is the hallmark of the residual gauge freedom (conformal symmetry). Then, taking the divergence of \eqref{wbg:Virasoro2} gives
\begin{equation}
\label{wbg:Virasoro2.div}
	\Box \zeta^{(2)\alpha} = -\nabla_\beta T^{\alpha\beta}~,
\end{equation}
which is solved by
\begin{equation}
\label{wbg:zeta2.sol}
	\zeta^{(2)\alpha} = -\frac1{\Box}\nabla_\beta T^{\alpha\beta} + \hat{\zeta}^{(2)\alpha}~,\qquad 
	\Box \hat{\zeta}^{(2)\alpha}=0~.
\end{equation}
Now we re-compose $\hat{\zeta}^\alpha = \zeta^{(1)\alpha} + \hat{\zeta}^{(2)\alpha}$, which is harmonic, 
\begin{equation}
\label{wbg:tilde.zeta.harmonic}
	\Box \hat{\zeta}^\alpha=0~,
\end{equation} 
and re-write \eqref{wbg:Virasoro2} in terms of $\hat{\zeta}^\alpha$. This yields
\begin{align}
\label{wbg:Virasoro3}
	2\nabla_{(\alpha}\hat{\zeta}_{\beta)} - g_{\alpha\beta} \nabla_\gamma \hat{\zeta}^\gamma 
	+ (\nabla_\alpha \hat{\zeta}_\gamma)(\nabla_\beta \hat{\zeta}^\gamma) - \frac12 g_{\alpha\beta} (\nabla_\gamma \hat{\zeta}_\delta)(\nabla^\gamma \hat{\zeta}^\delta) 
	=-\hat{T}_{\alpha\beta}~,
\end{align}
where, using \eqref{wbg:T.tilde},
\begin{equation}
\label{wbg:T.hat}
	\hat{T}_{\alpha\beta} 
	= \tilde{T}_{\alpha\beta} + \left( \frac{\nabla_\alpha\nabla_\beta}{\Box} -g_{\alpha\beta} \right) \tilde{T}~.
\end{equation}
Both sides of \eqref{wbg:Virasoro3} are now evidently conserved and traceless. In this way, we have re-written the Virasoro constraint as a constraint on an otherwise free harmonic vector, just as in the flat spacetime case.
The explicit calculation of $\hat{T}_{\alpha\beta}$ is deferred to appendix~\ref{app:T}. 

We proceed with the mode expansion of \eqref{wbg:Virasoro3}. This is easiest in chiral coordinates, $\sigma^\pm=\tau\pm \sigma$. On the one hand, because $\hat{T}_{\alpha\beta}$ is conserved and traceless, its two independent components are simply
\begin{equation}
\label{wbg:Tmmpp}
	\hat{T}_{--} = \alpha' \sum\limits_n L_n \e{-in\sigma^-}~,\qquad 
	\hat{T}_{++} = \alpha' \sum\limits_n \tilde{L}_n \e{-in\sigma^+}~,
\end{equation} 
with the Virasoro coefficients given by \eqref{appT:L} for $n\neq0$ and by \eqref{appT:L0.full} and \eqref{appT:tildeL0.full} for $n=0$. On the other hand, $\hat{\zeta}^\alpha$ is harmonic, which implies that its mode expansion reads
\begin{equation}
\label{wbg:zeta.expand}
	\hat{\zeta}^\alpha = \kappa^\alpha(\sigma^+ +\sigma^-) + \sum\limits_{n\neq0} \frac{i}n 
	\left( \zeta^\alpha_n \e{-in\sigma^-} + \tilde{\zeta}^\alpha_n \e{-in\sigma^+} \right)~.
\end{equation}
The zero mode coefficients $\kappa^\alpha$ are real, whereas the others satisfy $\zeta^\alpha_{-n}=\zeta^\alpha_{n}{}^\ast$, and similarly $\tilde{\zeta}^\alpha_{-n}=\tilde{\zeta}^\alpha_{n}{}^\ast$. 
Note that we can safely drop a constant vector and we have done so, because $\hat{\zeta}^\alpha$ appears in \eqref{wbg:Virasoro3} only within derivatives. Moreover, the only allowed linear zero mode respecting the periodicity of the closed string is the one proportional to $\tau$.

Inserting \eqref{wbg:Tmmpp} and \eqref{wbg:zeta.expand} into \eqref{wbg:Virasoro3} then yields the set of equations 
\begin{subequations}
\label{wbg:Vir.modes}
\begin{align}
\label{wbg:Vir.L0}
	\kappa^+(1+\kappa^-) +\sum\limits_{n\neq0} \zeta^+_n \zeta^-_{-n} - \frac{6}{\rho^2 \lambda} L_0 &=0~,\\
\label{wbg:Vir.R0}
	\kappa^-(1+\kappa^+) +\sum\limits_{n\neq0} \tilde{\zeta}^+_n \tilde{\zeta}^-_{-n} - \frac{6}{\rho^2 \lambda} \tilde{L}_0 &=0~,\\
\label{wbg:Vir.Ln}
	(1+\kappa^-)\zeta_n^+ + \kappa^+ \zeta_n^- +\sum\limits_{k\neq0}\sum\limits_{l\neq0} \delta_{k+l,n} \zeta_k^+ \zeta_l^-
	- \frac{6}{\rho^2 \lambda} L_n &=0~,\\
\label{wbg:Vir.Rn}
	(1+\kappa^+)\tilde{\zeta}_n^- + \kappa^- \tilde{\zeta}_n^+ +\sum\limits_{k\neq0}\sum\limits_{l\neq0} \delta_{k+l,n} 
	\tilde{\zeta}_k^+ \tilde{\zeta}_l^- - \frac{6}{\rho^2 \lambda} \tilde{L}_n &=0~,
\end{align}
\end{subequations}
where we used the relation $m_0 R^3/\alpha' = \lambda/6$. To continue, we note that there still is the residual gauge freedom corresponding to conformal symmetry, although this gauge freedom is non-linearly realized. We can use the residual symmetry to eliminate half of the oscillator modes in $\hat{\zeta}^\alpha$. More precisely, because a conformal Killing vector $k^\alpha$ satisfies $\partial_+ k^- =0$ and $\partial_- k^+=0$, the modes with $\zeta^-_n$ and $\tilde{\zeta}^+_n$ in \eqref{wbg:zeta.expand} represent conformal Killing vectors, and we can set these coefficients to zero.\footnote{Alternatively, on may set $\zeta^+_n=\tilde{\zeta}^+_n=0$, which is akin to fixing light-cone gauge.}
This simplifies the system \eqref{wbg:Vir.modes} to 
\begin{subequations}
\label{wbg:Vir.modes.simple}
\begin{align}
\label{wbg:Vir.simp.L0}
	\kappa^+(1+\kappa^-) - \frac{6}{\rho^2 \lambda} L_0 &=0~,\\
\label{wbg:Vir.simp.R0}
	\kappa^-(1+\kappa^+) - \frac{6}{\rho^2 \lambda} \tilde{L}_0 &=0~,\\
\label{wbg:Vir.simp.Ln}
	(1+\kappa^-)\zeta_n^+ - \frac{6}{\rho^2 \lambda} L_n &=0~,\\
\label{wbg:Vir.simp.Rn}
	(1+\kappa^+)\tilde{\zeta}_n^- - \frac{6}{\rho^2 \lambda} \tilde{L}_n &=0~.
\end{align}
\end{subequations}

\subsection{Ground state in flat spacetime}
\label{cov:hagedorn.flat}

As an intermezzo, let us verify that the geometrical formalism reproduces the known results for the mass-shell condition and the Hagedorn temperature in flat spacetime. In flat spacetime, the semiclassical approach is exact, because the absence of curvature implies that there are no terms of order higher than two in the action. 
Let us consider the case of flat spacetime with a compact Euclidean time circle, $x^1\sim x^1+\beta$, and a background string worldsheet given by
\begin{equation}
\label{intro:embed} 
	x^0 = \frac{\beta}{2\pi} \tau~, \qquad x^1 = \frac{\beta}{2\pi} \sigma~.
\end{equation}
Its induced metric is
\begin{equation}
\label{intro:induced.metric}
	\rmd s^2 = g_{\alpha\beta}\rmd \sigma^\alpha \rmd \sigma^\beta = \left(\frac{\beta}{2\pi}\right)^2 \left( -\rmd\tau^2 +\rmd \sigma^2\right)~.
\end{equation}
In the Virasoro constraint \eqref{wbg:Virasoro}, the tensors $T^{B}_{\alpha\beta}$ and $T^{F}_{\alpha\beta}$ are the 2-d stress-energy tensors for eight massless scalars and eight fermions, respectively, with suitable normalization. These are standard conformal field theory expressions familiar in string theory. Being quadratic in the fields, $T^{B}_{\alpha\beta}$ and $T^{F}_{\alpha\beta}$ contain normal-ordering constants, which determine the ground state configuration. More precisely,
\begin{align}
\label{intro:TB}
	T^{B}_{\tau\tau} &= \;:T^{B}_{\tau\tau}: +  8 \alpha' \sum\limits_{n=1}^\infty n = \;:T^{B}_{\tau\tau}: - 8\frac{\alpha'}{12}~,\\
\label{intro:TF}	
	T^{F}_{\tau\tau} &= \;:T^{F}_{\tau\tau}:- 8 \alpha' \sum\limits_{n=0}^\infty \left(n+\frac12\right) = \;:T^{F}_{\tau\tau}:- 8\frac{\alpha'}{24}~,
\end{align}
while 
\begin{equation}
\label{intro:Tts}
	T^{B}_{\tau\sigma}=\;:T^{B}_{\tau\sigma}:~,\qquad T^{F}_{\tau\sigma}=\,:T^{F}_{\tau\sigma}:~.
\end{equation}
The normal-ordered parts in \eqref{intro:TB}, \eqref{intro:TF}, and \eqref{intro:Tts} contain both constant as well as left- and right-moving Virasoro modes. However, in the ground state these terms vanish, so that the ground state solution of \eqref{wbg:Virasoro} has the form $\zeta^\tau = \kappa \tau$ with $\kappa$ a constant, and $\zeta^\sigma=0$. The quadratic equation for $\kappa$ that ensues is 
\be \label{intro:eq.for.k}
\kappa^2 + 2\kappa + \frac{8\pi^2\alpha'}{\beta^2} = 0\,.
\ee
Notice that the total energy of the string is
\be
p^0 = \frac{1}{2\pi\alpha'}\int d\sigma\, \partial_{\tau} x^0 = \frac{\beta}{2\pi\alpha'}(1+\kappa)\,.
\label{relkm}
\ee
Using the above relation, we see that eq. \eqref{intro:eq.for.k} is equivalent to the flat-space mass-shell condition on the ground state, that is
\be
\alpha' M^2 = \frac{\beta^2}{4\pi^2\alpha'}-2\,,
\label{massflat}
\ee
since (in the center of mass frame)
\be
M^2 = {(p^0)}^2 = \frac{\beta^2}{4\pi^2{\alpha'}^2}(1+\kappa)^2\,.
\ee
In flat space \eqref{intro:eq.for.k} or \eqref{massflat} are actually exact relations at the quantum level. The quadratic equation \eqref{intro:eq.for.k} is solved by
\be 
\label{cov:kappa.sol}
\kappa = - \left( 1\pm \sqrt{ 1-\frac{8\pi^2\alpha'}{\beta^2}} \right)~.
\ee 
This clearly shows that $\beta \geq \sqrt{8\alpha'}\pi$, which implies the existence of a limiting (Hagedorn) temperature $T_H = \frac1{\sqrt{8\alpha'}\pi}$.\footnote{In the case of the bosonic string, one would have $T_{\alpha\beta}$ for 24 massless scalars, giving a normal ordering constant of $-2\alpha'$ to $T^{B}_{\tau\tau}$. The same argument then leads to the Hagedorn temperature $T_H = (4\pi\sqrt{\alpha'})^{-1}$.} Equivalently, from \eqref{massflat}, this is the temperature at which the winding mode ground state becomes massless \cite{Atick:1988si}.

\subsection{Ground state in WYM}
\label{cov:hagedorn}

Let us return to our main argument and calculate the mass-shell condition in WYM. We shall follow the argument given in the preceding subsection and derive it directly from the constraint \eqref{wbg:Virasoro3}. We may also have started from the mode equations \eqref{wbg:Vir.modes.simple}, but this is not necessary for the ground state. In the vacuum, the normal-ordered products of the mode operators in $\hat{T}_{\alpha\beta}$ vanish, so that the vacuum expectation value coincides with the sum of the normal-ordering constants. Summing the contributions of all fields (see appendix~\ref{app:T}), we have
\begin{align} 
\label{wbg:T.hat.vac1}
	\braket{0|\hat{T}_{\tau\tau}|0} &= \alpha' \left[ \mu + \sum\limits_{n>0} \left(6 n + 2 \omega_n\right) -8\sum\limits_{r>0} \omega_r\right] =  -\alpha' \Delta(\mu)~,\\
\label{wbg:T.hat.vac2}
	\braket{0|\hat{T}_{\tau\sigma}|0} &=0~.
\end{align}
In \eqref{wbg:T.hat.vac1}, the explicit $\mu$ is the $n=0$ term from the massive scalars. The last equality in \eqref{wbg:T.hat.vac1} defines the same function $\Delta(\mu)$ as in \eqref{Delta:D.def}. The calculation of $\Delta(\mu)$ is carried out in appendix~\ref{Delta}. 

As there are no left- or right-moving modes in \eqref{wbg:T.hat.vac1}, we can set $\hat{\zeta}^\tau = \kappa \tau$, $\hat{\zeta}^\sigma=0$, which is \eqref{wbg:zeta.expand} without oscillators and with $\kappa^+=\kappa^- = \frac12 \kappa$. The non-trivial equation that arises from the $\tau\tau$ component of \eqref{wbg:Virasoro3} is\footnote{Equation~\eqref{wbg:lambda.eq} is just the sum of \eqref{wbg:Vir.simp.L0} and \eqref{wbg:Vir.simp.R0} with $L_0=\tilde{L}_0 = -\frac12\Delta(\mu)$.} 
\begin{equation}
\label{wbg:lambda.eq}
	-\frac{\lambda}{6}\rho^2 \left( \kappa +\frac12 \kappa^2\right)  - \Delta(\mu) =0~.
\end{equation}
The map between the two approaches is provided by the calculation of the total energy of the string
\be
p^0 = \frac{1}{2\pi} \int d\sigma\, \partial_{\tau} x^0\,,
\ee
which, for the geometrical approach, gives
\be
M \equiv p^0 = \rho \, (1+\kappa) \,.
\label{relkm}
\ee
Notice that the oscillators do not contribute to the integral as they are harmonic functions. Inserting the above relation in \eqref{wbg:lambda.eq} we get the mass-shell condition \eqref{massshellcond} evaluated on the ground state.

\section{Hagedorn temperature}
\label{TH}
As we have already recalled, at the Hagedorn temperature the ground state of the winding string becomes massless, $M^2=0$. Equivalently, according to \eqref{relkm}, this limit would be achieved when the amplitude of the longitudinal zero mode in the geometrical approach is $\kappa=-1$. This is not parametrically small, but leads to a longitudinal shift that is of the same order as the background. At the same time, the background parameter $\rho$ (related to the inverse temperature) becomes of order $1/\sqrt{\lambda}$, \ie parametrically small, in the Hagedorn regime.  
These observations force us to conclude that the Hagedorn regime cannot be consistently captured by the semiclassical approach, at least not in the form we have considered in the previous sections, because it rests on the assumption that there is a parametric hierarchy between the classical background and the quantum fluctuations.

In this section, we focus on the near-Hagedorn regime. In particular, in section~\ref{extoHag}, we propose how to adapt the results of the worldsheet perturbative approach of the previous sections to this regime, which leads to a prediction of the Hagedorn temperature of the WYM theory at sub-leading order in $1/\sqrt{\lambda}$. Then, in section~\ref{complapp}, we compare this result with the outcome of an effective model considered in a recent work by Urbach \cite{Urbach:2023npi} %(and to his result), 
finding perfect agreement. This will also provide a worldsheet interpretation of the results obtained from the effective approach.

\subsection{Worldsheet approach in the near-Hagedorn regime}
\label{extoHag}
To begin with, let us take into account that, in the Hagedorn regime, the background parameters $M$ and $\rho$ and thus also the mass parameter $\mu$ are of the same small order $\sim \lambda^{-1/2}$ as (some of\footnote{See later in footnote \ref{zeromodescaling}.}) the fluctuations. This places $M$ and $\rho$ in the mass-shell condition \eqref{massshellcond} at the same order as the contribution from the normal-ordering term with $\Delta(\mu)$.
To lowest order, at which we can neglect $\mu$, \eqref{massshellcond} reads
\be
M^2 = \rho^2 -\frac{12}{\lambda}\Delta(0) = \rho^2 -\frac{12}{\lambda}\,.
\label{nms0}
\ee
Here, in fact, $\Delta(0)=1$ is the normal ordering contribution due to 8+8 massless worldsheet scalars and fermions.
Setting $M^2=0$ we get the leading-order WYM Hagedorn temperature
\be
\rho_H=\sqrt{\frac{12}{\lambda}}\,,\quad {\rm i.e.}\,\, \frac{1}{T_H} \equiv \beta_H = \sqrt{\frac{4\pi}{T_s}}\,,
\label{thlead}
\ee
where
\be
T_s = \frac{m_0^2 \lambda}{12\pi}\,,
\ee 
is the confining string tension of the WYM model. 
This result, already claimed in \cite{Bigazzi:2022gal}, confirms our expectation, as $\rho_H\sim \lambda^{-1/2}$ is actually parametrically small. Notice also that the relation $\beta_H=\sqrt{4\pi/T_s}$ holds exactly in the flat-space case \eqref{cov:kappa.sol}. In the WYM model, as we are going to show below, this relation receives corrections which, to next-to-leading order at strong coupling, are captured by the subleading term in the expansion of $\Delta(\mu)$ around $\mu\ll1$.
%Therefore, we are confident that the modification needed in order for the worldsheet calculation to retain its validity is to incorporate a quantum correction for the mass parameter $\mu$.

In the semiclassical approach of the previous sections, the classical Virasoro constraint was given by $M^2=\rho^2$: imposing this relation, the mass parameter \eqref{mugen} became \eqref{mu.rho}. In the Hagedorn regime, instead,  \eqref{nms0} constitutes the Virasoro constraint at leading order; %\footnote{The identification of the background parameter $M$ with the mass of the physical string is motivated by the fact that the longitudinal fluctuations can be set to zero in the static gauge \cite{Forini:2015mca}. Static gauge would require to deal with the fluctuations of the intrinsic metric, the functional integration of which is subtle.} 
substituting this relation into \eqref{mugen} yields the (quantum corrected) value of the dimensionless mass parameter $\mu$; near the Hagedorn temperature (\ref{thlead}), it thus reads
\be
\label{TH:mu.corrected}
\mu^2 = \frac{9}{8}(M^2 + \rho^2) \rightarrow \frac{9}{8} \rho^2_H \sim \lambda^{-1}\,.
\ee
Then, keeping in \eqref{massshellcond} the next-to-leading term $\sim \mu$ in $\Delta(\mu)$ [see \eqref{Delta:D.result}], which is entirely due to the normal ordering contribution of the zero modes of the two massive bosons,\footnote{\label{zeromodescaling}The bosonic contribution to the mass shell condition is encoded by the canonical Hamiltonian as in \eqref{HamB}. If the scalar fields $y^i$ are taken to scale as $\lambda^{-1/2}$, in the near Hagedorn regime the mass terms of their non-zero modes $\mu^2 y^i_n y^i_n$ are of ${\cal O}(\lambda^{-2})$. This is not the case for the zero modes: from the structure of the mode expansion in \eqref{wbg:scal.mode.massive}, we see that the zero modes $y^i_0$ get a further scaling with $\mu^{-1/2}$. Hence, their Hamiltonian, i.e. $(\partial_{\tau}y^i_0)^2 + \mu^2 y^i_0 y^i_0$, is of ${\cal O}(\lambda^{-3/2})$.} we get
\be
M^2 = \rho^2 -\frac{12}{\lambda}(1- \mu)= \rho^2 -\frac{12}{\lambda}\left(1-\frac{3}{2\sqrt{2}}\sqrt{M^2 + \rho^2}\right)\,.
\label{nms1}
\ee
Setting again $M^2=0$, this gives
\begin{equation}
\label{wbg:rho.H.expand}
	\rho_H = \sqrt{\frac{12}{\lambda}} - \frac{1}{\sqrt{2}} \, \frac{9}{\lambda} + \mathcal{O}(\lambda^{-\frac32})\,,
\end{equation}
so that, to next-to-leading order in the strong coupling expansion, we get
\be
\frac{1}{T_H}\equiv \beta_H = \sqrt{\frac{4\pi}{T_s}} - \frac{1}{\sqrt{2}} \frac{M_{\mathrm{KK}}}{T_{s}} + O(M_{KK}^2 T_s^{-3/2})\,,
\label{thapp}
\ee
where $M_{\mathrm{KK}} =\frac32 m_0$, as discussed in section~\ref{background}. 

Going further in the perturbative expansion would require to take into account other corrections that we are neglecting. For example, next-to-next-to-leading order corrections to $\beta_H$ should be affected by one-loop corrections and ($\alpha'/R^2$)-corrections to the supergravity solution.\footnote{In \cite{Bigazzi:2004ze}, it was pointed out that the one-loop corrections produce an extra $\mathcal{O}(\lambda^{-1})$-term in the effective string tension. Moreover, the corrections to the supergravity background come from the Lagrangian term $\alpha'^3 \mathcal{R}^4$. Therefore, the background is typically affected at order $\mathcal{O}(\lambda^{-3/2})$ \cite{Buchel:2009nx}. All these contributions do not affect the first sub-leading correction to $\beta_H$.} However, the next-to-leading order $\mathcal{O}(\lambda^{-1})$-term in our outcome for the Hagedorn temperature in \eqref{thapp} is robust. 

Notice that if we had blindly started from the mass-shell conditions \eqref{massshellcond} or \eqref{wbg:lambda.eq} with the classical relation $\mu=\frac32\rho$, expanding $\Delta(\mu)=1-\mu +{\cal O}(\mu^2)$, we would have found a result analogous to  \eqref{thapp}, but with a coefficient $1$ instead of $1/\sqrt{2}$ in front of the sub-leading term. This is a consequence of the fact that the quantum corrected \eqref{TH:mu.corrected} differs by a factor of $\frac12$ from the classical relation at the Hagedorn temperature.

%%%%%%%%%%%%%%%%%%%%%%%%%%
\subsection{Effective string gas model}
\label{complapp}

In a very recent paper \cite{Urbach:2023npi}, Urbach has calculated the Hagedorn temperature for the WYM theory (and other holographic confining models) to next-to-leading order in the holographic limit, using an effective approach inspired by the Horowitz-Polchinski construction \cite{Horowitz:1997jc} (see also \cite{Chen:2021dsw}). The construction is particularly suited for the near-Hagedorn regime, where the temporal winding modes of the string become light. The approach has been previously used by the same author to compute the Hagedorn temperature of $\mathcal{N}=4$ super Yang Mills on $S^3$ \cite{Urbach:2022xzw}, finding a result in agreement, to next-to-leading order in the holographic limit, with the result obtained using integrability and quantum spectral curve methods \cite{Harmark:2017yrv, Harmark:2018red, Harmark:2021qma}.

Let us re-consider the computation in \cite{Urbach:2023npi} making explicit some details of the background (the non-trivial dilaton profile, for instance) and allowing for a non-zero momentum of the winding modes. The starting point is the type-IIA supergravity action on the near-tip WYM background \eqref{Wbacktipy}, with the addition of the complex scalar $\chi$ corresponding, as in the flat space case \cite{Atick:1988si}, to the string states with winding numbers $\pm1$ on the temporal circle. The Hagedorn temperature of the theory is given by the temperature for which, neglecting the backreaction on the background, the linearized equation of motion for $\chi$ admits a normalizable solution. The solution turns out to be parametrically small, which a-posteriori justifies why we can work with just a quadratic effective action for the scalar field $\chi$.

Including the backreaction would actually allow to infer that the system undergoes a first order transition at a critical temperature $T_c<T_H$, where a small black-hole branch is expected to become energetically preferred. The backreaction is accounted for allowing for a fluctuation of the background metric of the form $G_{11}=g_{11}e^{\varphi}$. As in \cite{Urbach:2023npi}, turning off the contribution of $\varphi$ to the linearized equation of motion for $\chi$ corresponds to following the aforementioned metastable branch. In this way, we can circumvent the critical point at $T=T_c$ in order to probe the near Hagedorn regime.

After reduction on $S^4$, the effective action for $\chi$, up to $\mathcal{O}(\lambda^{-1})$ corrections, is expected to read
\be
S_{\chi} \sim \int d^4 x\, d^2 y \sqrt{\det g}\, e^{-2\Phi} \left[ g^{MN}\partial_M \chi \partial_N\bar \chi + \text{m}^2_{\text{eff}}(y) \bar\chi \chi \right]\,,
\ee
where\footnote{Here we follow the conventions of the previous sections and call $x^1$ the compact Euclidean time direction.}
\be
\alpha' \, \text{m}^2_{\text{eff}}(y) = \frac{\beta^2 g_{11}(y) - 8\pi^2\alpha'}{4\pi^2\alpha'}=\rho^2 \, \frac\lambda6 \(1+\frac{9}{8}y^2\) - 2\,.
\ee
The leading order Hagedorn temperature is implied by $\text{m}^2_{\text{eff}}(0)=0$: this is the temperature at which the proper length of the time circle at the tip of the background equates the critical length $\beta_{H,\rm{flat}}= 2\pi\sqrt{2 \alpha'}$ in flat spacetime.  

Neglecting the backreaction of $\chi$ on the geometry and using the ansatz\footnote{Here, the `t Hooft coupling appears in the phase of the plane wave since the canonical momenta $p_c^i$ are defined as $p_c^i = T_s \int d\sigma \partial_\tau x^i/m_0$.}
\be
\chi(\vec x, y) = e^{i \, \frac\lambda6 \vec p \, \cdot \, \vec x} w(y) \, \quad \vec x = \{x^0, x^2, x^3\}~,
\ee
the equation of motion for $w$ reads
\be \label{eomw}
-\frac12 w''(y) - \frac12 \frac{w'(y)}{y} -\frac{9}{8} y\, w'(y) \left(1-\frac{9}{8}y^2\right) + \frac12 \, a \, y^2 w(y) = b \, w(y)
\ee
with the coefficients
\be
a= \frac98 \[ M^2 + \rho^2 \] \(\frac\lambda6\)^2 \, , \quad b = \frac12 \[  \(\frac\lambda6\)^2 (M^2 - \rho^2) + 2 \, \frac\lambda6\] \,,
\ee
and $M^2 =-\eta_{ij} p^ip^j$. If we now re-scale $y = \Lambda \tilde y$, with $\Lambda^{-4} = a$, we see that the third term on the left hand side of  (\ref{eomw}) can be neglected for $\lambda\gg1$, and the equation reduces to
\be \label{eomw2}
-\frac12 w''(\tilde y) - \frac12 \frac{w'(\tilde y)}{\tilde y} + \frac12 \tilde y^2 w(\tilde y) = \frac{b}{\sqrt{a}} \, w(\tilde y)\,.
\ee
Eqn.~\eqref{eomw2} admits a normalizable solution, namely $w(\tilde y) \sim e^{-\tilde y^2/2}$, if $b=\sqrt{a}$. This relation, in turn, translates into
\be \label{effmassshell}
\frac{\lambda}{12} (-M^2 + \rho^2)  = 1 - \frac{3}{2\sqrt{2}}\sqrt{M^2+\rho^2} \,,
\ee
which precisely matches with our mass-shell condition \eqref{nms1}. Hence, setting $M^2=0$, both methods give the same result for the Hagedorn temperature.
This matching allows us to outline that the next-to-leading correction to the Hagedorn temperature, as computed in \cite{Urbach:2023npi}, is due to the normal ordering of the zero modes of the two massive worldsheet scalar fields.\footnote{In \cite{Urbach:2023npi} the sub-leading correction to the Hagedorn temperature for a generic holographic confining theory dual to a background with a $d-1$-dimensional shrinking cycle, was found to be proportional to $d$. In the semiclassical string approach $d$ (with $d=2$ in the WYM case) is precisely the number of massive worldsheet scalar fields.}

Notice that when $M$ and $\rho$ are generic, $\Lambda^2\sim \lambda^{-1}$ and the size of the normalizable solution $w(\tilde y)$ is of the order of the string size, $\tilde y^2\sim r^2/l_s^2$, where
$r^2 =\lambda\, l_s^2\, y^2/6$ and $l_s^2=\alpha'$. Instead, when $M\rightarrow0$ and $\rho\rightarrow\rho_H$, $\Lambda^2\sim \lambda^{-1/2}$ and the size of the solution is much larger than the string size, since now $\tilde y^2\sim r^2/L^2$, where $L^2\sim l_s l$, with $l^2 = (8/9) \lambda\, l_s^2/6$. The latter is in fact the regime where the effective approach is meant to be sensible.

\section{Conclusions}
\label{conc}

We have carried out two different derivations of the mass-shell condition for a type IIA string winding once around the compact Euclidean time circle and placed at the tip of the cigar in the WYM background. The calculation in section~\ref{lightcone} makes rigorous the previous analysis of \cite{Bigazzi:2022gal}, where the role of the second-order corrections to the equations of motion and the need to relax the light-cone gauge choice to second order were missing. Readers wanting to see the rigorous argument in polar coordinates adopted in \cite{Bigazzi:2022gal} can consult appendix~\ref{appUT}.  
The new calculation in section~\ref{covariant} is based on a coordinate-invariant expansion of the string fluctuations around a winding background string. In this approach, the fluctuation fields possess a clear geometrical meaning. 

Both semiclassical calculations are consistent as long as the main assumption of perturbation theory holds: the fluctuations  must be small compared to the typical background scale. In the case at hand, this is true for sufficiently low temperature, but not near the Hagedorn temperature, at which the ground state of a string winding once around the thermal compact direction is massless \cite{Atick:1988si}. Therefore, although the naive extrapolation of the semiclassical calculations also predicts a limiting temperature (see comments in the last paragraph in subsection~\ref{extoHag}), it does not agree with the actual Hagedorn temperature beyond leading order.\footnote{The result is correct in Minkowski spacetime, because there the semiclassical calculation is exact, see subsection~\ref{cov:hagedorn.flat}.}

In section \ref{TH}, we have thus adopted a different perturbative scheme, where the leading order ground state mass-shell condition is given by \eqref{massshellcond}. 
This affects the dimensionless mass parameter $\mu$ as in \eqref{TH:mu.corrected} and effectively permits to extend the semiclassical formulae to the near-Hagedorn regime. In turn, this has allowed us to compute the Hagedorn temperature of WYM theory at next-to-leading order in $1/\sqrt{\lambda}$, see eq. \eqref{thapp}. We have also verified this result with an effective action approach based on the Horowitz-Polchinski string star model and the recent work \cite{Urbach:2023npi}. The worldsheet perspective presented in our work allows to give an interpretation to the next-to-leading order correction, as that due to the normal ordering of the zero modes of the massive scalar worldsheet fields. This observation holds for general holographic confining models, as those considered in \cite{Urbach:2023npi}. The dual background metrics exhibit a shrinking $(d-1)$-cycle, and the sub-leading correction to the Hagedorn temperature is due to the related $d$ massive worldsheet scalar fields ($d=2$ for the WYM theory).

It is an interesting observation that the quantum-corrected expression for $\mu$ agrees with \eqref{mugen} without imposing the Virasoro constraint on the background worldsheet. The question that arises is whether the semiclassical calculation could have been erected on a background, which does not satisfy the classical Virasoro constraint. 
This change of scheme has profound implications, especially within the geometrical method where, without the Virasoro constraint on the background worldsheet, one loses background covariance, because the auxiliary metric is not conformal to the background induced metric. Moreover, the tangential fluctuations become dynamical fields, although they remain constrained. 
We hope to return to these issues in the future. 

%%%%%%%%%%%%%%%%%%%%%%%%%%%%%%%%%%%%%%%%%%%%%%%%%%%%%%%%%%%%%%%%%%%%%%%%%%%%%%%%%%%%%%%%%%%%%%%%%%%%%%%%%%%%%%%%%%%%%%%%%

\section*{Acknowledgments}
We are greatly indebted to A.L.~Cotrone for many comments, discussions and suggestions. We thank Y.~Chen, L.~Martucci and D.~Seminara for comments.
We acknowledge support by the INFN research initiatives GAST and STEFI.

\begin{appendix}
\section{Geometry of embedded manifolds}
\label{embed}
\subsection{Structure Equations}
\label{embed:structure}
The differential geometry of embedded manifolds has been described in
detail by Eisenhart \cite{Eisenhart}. 
Consider a $d$-dimensional (pseudo-)Riemannian manifold $\mathcal{M}$ embedded in a $D$-dimensional (pseudo-)Riemannian manifold $\mathcal{B}$. $\mathcal{M}$ and $\mathcal{B}$ will be called the \emph{worldsheet} and the \emph{bulk}, respectively, with an eye to an application to strings. 

Let $x^m$ and $g_{mn}$, $m,n=1,2,\ldots D$, be a set of coordinates and the corresponding metric in the bulk.  
The coordinates on the worldsheet are $\sigma^\alpha$, $\alpha,\beta=1,2,\ldots, d$. 
The embedding of $\mathcal{M}$ in $\mathcal{B}$ is locally described by $D$ differentiable functions $x^m(\sigma)$ specifying point-wise the position of $\mathcal{M}$ within $\mathcal{B}$. 

The tangent vectors on $\mathcal{M}$ are 
\begin{equation}
\label{embed:tangents}
  x^m_\alpha(\sigma) \equiv \partial_\alpha x^m(\sigma)~.
\end{equation}
The tangents define the pull-back of tensors from $\mathcal{B}$ to $\mathcal{M}$. In particular, the induced metric is 
\begin{equation}
\label{embed:g}
  g_{\alpha\beta} = x^m_\alpha x^n_\beta g_{mn}~. 
\end{equation}

To get a locally complete basis of vectors in the bulk, introduce $D-d$ normal vectors, $N_i^m$, $i,j= 1,2,\ldots , D-d$.  
They may be chosen to satisfy the orthogonality relations
\begin{subequations}
\label{embed:ortho}
\begin{align}
  N_i^m x_\alpha^n g_{mn} &=0 ,\\
  N_i^m N_j^n g_{mn} &= \eta_{ij}~,
\end{align} 
\end{subequations}
where $\eta_{ij}$ is a (pseudo-)Euclidean metric, which takes into account the signatures of $\mathcal{B}$ and $\mathcal{M}$. Most often, it is Euclidean. The vectors $N_i^m$ are said to span the \emph{normal bundle}. The normal bundle possesses a (pseudo-)$O(n)$ symmetry implementing the freedom to rotate the normals. This symmetry is gauged on the worldsheet. Moreover, the normals $N^m_i$ have an obvious interpretation as the part of a vector frame that is transverse to $\mathcal{M}$ ($N^m_i = e^m_i$).  
Completeness of the vector basis means 
\begin{equation}
\label{embed:complete}
	g^{\alpha\beta} x_\alpha^m x_\beta^n + \eta^{ij} N_i^m N_j^n = g^{mn}~.
\end{equation}

The embedding is characterized by a number of structure equations. These are, first, the equation of Gauss, 
\begin{equation}
\label{embed:gauss1}
  \nabla_\alpha x^m_\beta \equiv \partial_\alpha x^m_\beta +
  \Gamma^m{}_{np} x^n_\alpha x^p_\beta - \Gamma^\gamma{}_{\alpha\beta} 
  x^m_\gamma = H^i{}_{\alpha\beta}N_i^m~,
\end{equation}
which defines the \emph{second fundamental forms}, $H^i{}_{\alpha\beta}$. These describe the \emph{extrinsic curvature} of the embedding of $\mathcal{M}$ within $\mathcal{B}$.

Second, there is the equation of Weingarten, 
\begin{equation}
\label{embed:weingarten}
  \nabla_\alpha N^m_i \equiv \partial_\alpha N^m_i +
  \Gamma^m{}_{np} x^n_\alpha N^p_i - A^j{}_{i\alpha} N^m_j = -
  H_{i\alpha}{}^\beta x^m_\beta~.
\end{equation}
It introduces the \emph{connections in the normal bundle}, $A_{ij\alpha}$. They are antisymmetric, $A_{ij\alpha}=-A_{ji\alpha}$ and represent the gauge fields associated with the local (pseudo-)$O(n)$ symmetry.
The corresponding field strength is 
\begin{equation}
\label{embed:F}
  F_{ij\alpha\beta} = \partial_\alpha A_{ij\beta} - \partial_\beta
  A_{ij\alpha} + A_{ik\alpha} A^k{}_{j\beta} - A_{ik\beta}
  A^k{}_{j\alpha}~.
\end{equation}

Finally, the integrability conditions of the differential equations \eqref{embed:gauss1} and \eqref{embed:weingarten} are the equations of
Gauss, Codazzi and Ricci. These are, respectively,
\begin{align}
\label{embed:gauss2} 
  R_{mnpq} x^m_\alpha x^n_\beta x^p_\gamma x^q_\delta &=
  R_{\alpha\beta\gamma\delta} + H^i{}_{\alpha\delta} H_{i\beta\gamma}
  - H^i{}_{\alpha\gamma} H_{i\beta\delta}~,\\
\label{embed:codazzi}
  R_{mnpq} x^m_\alpha x^n_\beta N^p_i x^q_\gamma &=
  \nabla_\alpha H_{i\beta\gamma} - \nabla_\beta H_{i\alpha\gamma}~,\\
\label{embed:ricci}
  R_{mnpq} x^m_\alpha x^n_\beta N^p_i N^q_j &= F_{ij\alpha\beta}
  - H_{i\alpha}{}^\gamma H_{j\gamma\beta} + H_{i\beta}{}^\gamma H_{j\gamma\alpha}~.
\end{align}

\subsection{Spin connections}
For spinors, we need \emph{spin connections}. The bulk spin connections are defined in terms of a frame $e^m_\flati{m}$ satisfying 
\begin{eqnarray}
\label{embed:frame}
	e^m_\flati{m} e^n_\flati{n} g_{mn} = g_{\flati{mn}} \equiv \eta_{\flati{mn}}~.
\end{eqnarray}
The underlined indices are flat indices and appropriately contracted with the metric $\eta_{\flati{mn}}$. The bulk spin connections are 
\begin{equation}
\label{embed:omega}
	\omega_p{}^\flati{mn} = -e_q^\flati{n} \left( \partial_p e^{q\flati{m}} +\Gamma^q{}_{pn} e^{n\flati{m}} \right)~,
\end{equation}
Analogous relations hold for the worldsheet spin connections, which derive from a frame on the worldsheet, $e^\alpha_{\flati{\alpha}}$.  
As mentioned above, the normal vectors $N^m_i$ are naturally interpreted as $D-d$ basis vectors of a frame that is locally adapted to the worldsheet. The $d$ missing basis vectors must be constructed from the tangents. Thus, a bulk frame locally adapted to the worldsheet is 
\begin{equation}
\label{embed:local.frame}
	e^m_\flati{n} = 
	\begin{cases} 
		x^m_\alpha e^\alpha_\flati{\alpha} &\text{for $\flati{n}=\flati{\alpha}$,} \\
		N^m_i &\text{for $\flati{n}=i$.} 
	\end{cases}
\end{equation}

For \eqref{embed:local.frame} one can show using \eqref{embed:gauss1} and \eqref {embed:weingarten} that
\begin{equation}
\label{embed:pb1}
	x_\alpha^m \left( \partial_m e^q_\flati{n} +\Gamma^q{}_{mp} e^p_\flati{n} \right) = \begin{cases} 
		N^q_i H^i{}_{\alpha\beta} e^\beta_\flati{\alpha} 
		+ x_\beta^q e^{\beta\flati{\beta}} \omega_{\alpha\flati{\beta\alpha}}
		& \text{for $\flati{n}=\flati{\alpha}$,}\\
		-x^q_\beta H_{i\alpha}{}^\beta  + N^q_j A^j{}_{i\alpha}  
		&\text{for $\flati{n}=i$.} 
	\end{cases}
\end{equation}
This yields, together with \eqref{embed:omega}, the pull-back of the bulk spin connections onto the worldsheet,
\begin{equation}
\label{embed:pb.spin.conn}
	x_\alpha^m \omega_{m\flati{\alpha\beta}} = \omega_{\alpha\flati{\alpha\beta}}~,\qquad 
	x_\alpha^m \omega_{mi \flati{\alpha}} = H_{i\alpha\beta} e^\beta_\flati{\alpha}~,\qquad
	x_\alpha^m \omega_{mij} = A_{ij\alpha}~.
\end{equation}

As an application, consider the (standard) covariant derivatives for bulk spinors,  
\begin{equation}
\label{embed:D.alpha}
	\nabla_m \Psi = \left( \partial_m +\frac14 \omega_m{}^\flati{np} \Gamma_\flati{np} \right)\Psi~.
\end{equation}
Using \eqref{embed:pb.spin.conn}, the pull-back of this derivative on the worldsheet is 
\begin{equation}
\label{embed:D}
	x_\alpha^m \nabla_m \Psi= \left( \partial_\alpha  + \frac14 \omega_{\alpha\flati{\beta\gamma}} \Gamma^\flati{\beta\gamma} 
	+\frac12 H_{i\alpha\beta} \Gamma^i \Gamma^\beta 
	+\frac14 A_{ij\alpha} \Gamma^{ij}\right) \Psi~.
\end{equation}
\section{Geometry of fluctuations}
\label{exp}

\subsection{Exponential map}

In order to parameterize the fluctuations around a background worldsheet, it is very convenient to use a manifestly coordinate-invariant formalism, in which the parameters are geometric entities \cite{Alvarez-Gaume:1981exa}. Coordinate differences, such as $\delta x^m = x^m - \bar{x}^m$, are usually not suitable, because they behave non-covariantly under coordinate transformations. 
The correct procedure is based on the exponential map from a root point $\bar{x}$ to a target point $x$. The exponential map is  unique, if $x$ falls within a certain neighbourhood of $\bar{x}$. Indeed, a theorem \cite{Willmore} ensures that there exists a unique geodesic curve $x^n(s)$ connecting $x$ and $\bar{x}$. Without loss of generality, $s$ can be taken as an affine parameter normalized such that $x^n(0)= \bar{x}^n$ and $x^n(1)=x^n$. Defining the tangent vector along this geodesic at $\bar{x}$ as $Y^n$, 
\begin{equation}
\label{exp:Y.def}
	Y^n = \left.\frac{\rmd x^n(s)}{\rmd s}\right|_{s=0}~,
\end{equation}
it can be shown by induction that the target point $x^n=x^n(1)$ is given by 
\begin{equation}
\label{exp:X.exp.map}
	x^m = \sum\limits_{k=0}^\infty \frac{1}{k!} \left( \nabla_{n_1}\ldots \nabla_{n_k} x^{[m]}\right)_{\bar{x}} Y^{n_1}\ldots Y^{n_k}~.
\end{equation}
Here, with a gross abuse of notation, we have used the formal notation of ``covariant'' derivatives and indicated with the brackets around the coordinate index $[m]$ that this index is spared from receiving connections. (The coordinate index is not a tensor index.) Explicitly, one has 
\begin{subequations}
\label{exp:dy}
\begin{align}
\label{exp:dy1}
  \nabla_n x^{[m]} &= \partial_n x^m = \delta^m_n~,\\ 
\label{exp:dy2}
  \nabla_{n_1} \nabla_{n_2} x^{[m]} &= - \Gamma^p{}_{n_1n_2} \nabla_p x^{[m]} = - \Gamma^m{}_{n_1n_2}~,\\ 
\label{exp:dy3}
  \nabla_{n_1} \nabla_{n_2} \nabla_{n_3} x^{[m]} &= - \Gamma^m{}_{n_2n_3,n_1} + \Gamma^p{}_{n_1n_2} \Gamma^m{}_{pn_3} + \Gamma^p{}_{n_1n_3} \Gamma^m{}_{pn_2}~, \\
\notag &\ldots~
\end{align}
\end{subequations}

Formally, the exponential map \eqref{exp:X.exp.map} is denoted by
\begin{equation}
\label{exp:exp.map.formal}
	Y: \bar{x} \to x = \exp_{\bar{x}}(Y)~.
\end{equation}
For brevity, we shall henceforth drop the bar from the root point $\bar{x}$ and denote the exponential map, as well as expansions that follow from it, by 
\begin{equation}
\label{exp:exp.map.formal2}
	x \to \exp_{x}(Y)~.
\end{equation}

Riemann normal coordinates (RNCs) \cite{Petrov} help to simplify many of the necessary calculations. These simplifications derive from the properties listed in the following theorem 
\begin{theorem}
\label{th2}
\cite{Petrov} Let $U$ be a neighborhood of a given point, which is chosen as the origin of the coordinates in $U$.
Each of the following three properties is a necessary and sufficient condition for coordinates $y^n$ in $U$ to be RNCs:
\begin{subequations}
\begin{gather}
\label{riem:cond1}
  \text{The geodesics are described by } y^n(s) =\upsilon^n s
  \text{ with $\upsilon^n=const.$}\\ 
\label{riem:cond2}
  \quad \Gamma^m{}_{np}\, y^n y^p =0 \quad\forall y\in U~,\\
\label{riem:cond3}
  g_{mn} \,y^m = \left.g_{mn}\right|_{y=0} \,y^m \quad\forall x \in U~.
\end{gather}
\end{subequations}
\end{theorem}
We immediately recognize that the \emph{components} of the generating vector $Y^n$ of exponential maps are RNCs in a neighborhood of the root point $\bar{x}$. 
The simplifications mentioned above derive from the fact that, for RNCs and only for those, theorem~\ref{th2} implies\footnote{Parentheses around indices denote their symmetrization, \eg $A_{(\alpha\beta)}=\frac12(A_{\alpha\beta} + A_{\beta\alpha} )$.}  
\begin{equation}
\label{exp:riem.simp}
	\nabla_{(n_1}\ldots \nabla_{n_k)} y^{[m]} =0 \qquad \text{for $k\geq 2$ (RNCs only)}~.
\end{equation}

As an example, the expansion of a covariant bulk tensor of rank $k$ to second order in $y^m$ is given by \cite{Petrov}
\begin{equation}
\label{exp:A.expand}
	A_{n_1\cdots n_k} \to A_{n_1\cdots n_k} +A_{n_1\cdots n_k;m} y^m + 
	\frac12\left( A_{n_1\cdots n_k;mp} +\frac13 \sum\limits_{l=1}^k R^q{}_{mpn_l} A_{n_1\cdots q\cdots n_k} \right) y^my^p +\cdots~.	
\end{equation}
For the metric tensor, this implies
\begin{equation}
\label{exp:g.expand}
	g_{mn} \to g_{mn} -\frac13 R_{mpnq} y^p y^q +\cdots~.	
\end{equation}
We stress that these relations hold only in RNCs. In general coordinates, one would find terms with connections on the right hand side. The reason is that a tensor $A_{n_1\cdots n_k}(x)$ at a point $x$ cannot, in general, be expressed in terms of tensors at another point $\bar{x}$, but one may parallel transport the tensor $A_{n_1\cdots n_k}$ from $x$ to $\bar{x}$ in order to find a tensor equation. The parallel transport is trivial in RNCs, but would introduce connections in other coordinate systems. The only exception to this occurs for scalars. Therefore, as long as one is interested in spacetime scalars (which may contain contracted tensors or may be worldsheet tensors), one can use RNCs and then replace the RNCs $y^n$ with the vector components $Y^n$, and the result will be coordinate-invariant.

In the present context, one must be a bit more careful. The reason is that the RNCs are defined with reference to a single point $\bar{x}$. However, when we consider the fluctuation of an embedded manifold, we deal with a background manifold of root points $\bar{x}(\sigma)$, and the RNCs around one point do not coincide with the RNCs around another point. For our purposes, it suffices to illustrate this with the tangent vectors $x_\alpha^n$. In general coordinates, the expansion of the tangents is found by differentiating \eqref{exp:X.exp.map} with respect to $\sigma^\alpha$, 
\begin{align}
\label{exp:tang.expand.1}
	x_\alpha^m &\to x_\alpha^m + x_\alpha^l \left[ (\nabla_l \nabla_n x^{[m]}) Y^n + (\nabla_n x^{[m]}) \nabla_l Y^n \right] \\
\notag
	&\quad 
	+\frac12 x_\alpha^l \left[ (\nabla_l \nabla_p \nabla_q x^{[m]}) Y^p Y^q + (\nabla_p \nabla_q x^{[m]}) (\nabla_l Y^p)Y^q + Y^p (\nabla_l Y^q)  \right] +\cdots~.
\end{align}
In RNCs, this simplifies to 
\begin{equation}
\label{exp:tang.expand.RNC}
	x_\alpha^m \to x_\alpha^m + \nabla_\alpha Y^m -\frac13 R^m{}_{pnq} x_\alpha^n Y^p Y^q+\cdots\qquad \text{(RNCs)}~.
\end{equation}
The expansion of the worldsheet induced metric is then found by combining \eqref{exp:g.expand} with \eqref{exp:tang.expand.RNC},
\begin{equation}
\label{exp:g.induced.expand}
	g_{\alpha\beta} \to g_{\alpha\beta} + 2 x_{(\alpha}^n \nabla_{\beta)} Y_n + (\nabla_\alpha Y_n)(\nabla_\beta Y^n)
	- R_{mpnq} Y^pY^q x_\alpha^m x_\beta^n +\cdots~.	
\end{equation}
As mentioned above, because $g_{\alpha\beta}$ is a spacetime scalar, \eqref{exp:g.induced.expand} holds in any coordinates.

\subsection{Normal and tangential components}
\label{exp:string}

The spacetime vector $Y^n$ is not a nice geometric object in relation to the background worldsheet, because its target space is the tangent space of spacetime. In order to introduce nicer objects, we decompose $Y^n$ into tangential and normal components with respect to the background worldsheet,
\begin{equation}
\label{exp:Y.decomp}
	Y^n = x_\alpha^n \zeta^\alpha + N_i^n \chi^i~.
\end{equation}
The worldsheet scalars $\chi^i$ for the normal fluctuations are, in general, charged under the gauge group in the normal bundle. This is implicit in the use of the covariant derivatives $\nabla_\alpha$ introduced in appendix~\ref{embed:structure}. The worldsheet vector $\zeta^\alpha$ parameterizes the longitudinal fluctuations. We do not have the freedom to gauge-fix them to zero, although there is a certain amount of gauge freedom that will be fixed later.

Now we have all the machinery ready to carry out the expansion around a classical background string worldsheet. All one needs to do is to substitute \eqref{exp:g.induced.expand} in combination with \eqref{exp:Y.decomp} in both the action \eqref{str:action} and the Virasoro constraint \eqref{str:Virasoro}. Moreover, the auxiliary metric $h_{\alpha\beta}$ must be replaced with the background $g_{\alpha\beta}$ using Weyl invariance and the fact that the background satisfies \eqref{str:Virasoro}. 
We remark that this is absolutely independent of the conformal gauge choice. In fact, the conformal gauge would only conveniently restrict the choice of coordinates on the background worldsheet, but in principle any choice is allowed.  

Let us start by substituting \eqref{exp:Y.decomp} into \eqref{exp:g.induced.expand}. Using the structure equations of appendix~\ref{embed:structure}, one obtains 
\begin{align}
\label{exp:g.expand.decomp}
	g_{\alpha\beta} &\to g_{\alpha\beta} + 2 \nabla_{(\alpha}\zeta_{\beta)} - 2 H_{i\alpha\beta}\chi^i \\
\notag &\quad 
	+ (\nabla_\alpha \zeta_\gamma)(\nabla_\beta \zeta^\gamma) + \left( H_{i\alpha\beta}H^i{}_{\gamma\delta} -R_{\alpha\gamma\beta\delta} \right) \zeta^\gamma\zeta^\delta \\
\notag &\quad 
	+2 \nabla_{(\alpha} \left( \zeta^\gamma \chi^i H_{|i|\beta)\gamma} \right) -4 \chi^i H_{i\gamma(\alpha} \nabla_{\beta)} \zeta^\gamma
	-2 \chi^i \zeta^\gamma \nabla_\gamma H_{i\alpha\beta} \\
\notag &\quad
	+ (\nabla_\alpha \chi^i)(\nabla_\beta \chi_i) + \left( H_{i\alpha\gamma} H_{j\beta}{}^\gamma -R_{mpnq}x_\alpha^m x_\beta^n N_i^p N_j^q \right) \chi^i \chi^j +\cdots~.
\end{align}

After substituting \eqref{exp:g.expand.decomp} into the bosonic action \eqref{str:Polyakov}, the first-order terms tell us that  the classical field equation for the background is 
\begin{equation}
\label{exp:H.field.equation}
	H_{i\alpha}{}^\alpha =0~.
\end{equation}
Let us proceed with the second order terms in the action \eqref{str:Polyakov}. We make use of the identity 
\begin{equation}
\label{exp:R.ident}
	R_{\alpha\gamma\beta\delta} \zeta^\gamma \zeta^\delta = \zeta^\gamma \left(\nabla_{(\alpha} \nabla_{|\gamma|} - \nabla_\gamma \nabla_{(\alpha}\right) \zeta_{\beta)}
\end{equation}
and the fact that the first-order terms in the Virasoro constraint are\footnote{We have used \eqref{exp:H.field.equation}.}
\begin{equation}
\label{exp:Vir.first}
	2 \nabla_{(\alpha}\zeta_{\beta)} -g_{\alpha\beta} \nabla_\gamma \zeta^\gamma - 2 H_{i\alpha\beta}\chi^i +\cdots =0~,
\end{equation}
which can be used to eliminate $\zeta^\alpha$ from the second-order terms of the action, except for some total derivative term. Dropping this irrelevant term, one obtains the action \eqref{str:S.P.expand}. 

It is not necessary, at the order we are working at, to expand $g_{\alpha\beta}$ in the fermion action \eqref{str:ferm.action}, where the background term is sufficient. Moreover, the Virasoro constraint \eqref{str:Virasoro} to second order involves the traceless part of \eqref{exp:g.expand.decomp}, but we refrain from writing it out here.

\section{Polar coordinates}
\label{appUT}
In section \ref{lightcone}, we adopt locally cartesian coordinates $y_1$ and $y_2$ around the tip of the cigar instead of the polar coordinates $u = u_0 \, v$ and $\theta$, as in \cite{Bigazzi:2022gal}. Of course, covariance guarantees that the final result does not depend on the choice of coordinates. And indeed that is what happens. In this appendix we briefly review the computation of section \ref{lightcone} in polar coordinates, in order to see this equivalence in a clear way.

As shown in \cite{Bigazzi:2022gal}, a consistent classical configuration in polar coordinates is the $U\to u_0$ limit of
\begin{subequations} \label{ansatz}
  \begin{align}
    &x^1 =\rho \sigma \, , \quad x^1 \simeq x^1 + 2\pi\rho \, ,\\
    &\label{Xi}x^i =p^i \tau \, , \quad p^i \in \mathbb{R} \quad : \quad - \eta_{ij} \, p^i \, p^j = M^2 \, , \quad i,j=0,2,3 \, ,\\
    &\label{firstU} u=U(\tau,\sigma) \quad : \quad \partial_\alpha U = \frac{c_\alpha}{g_{uu}(U)} \, , \quad c_\tau \, , c_\sigma \in \mathbb{R} \, , \\
    &\label{Theta} \Theta = 0 \, .
  \end{align}
\end{subequations}
Let us parametrize the fluctuations along the cigar directions as\footnote{The normalization of the first order fluctuations is required for them to have canonical kinetic terms in $S_B$.}
\be \label{cigarfluc}
u = U + \frac{1}{\sqrt{g_{uu}(U)}} \, \xi^u_{(1)}  + \xi^u_{(2)} \, \quad \theta = \frac{1}{\sqrt{g_{\theta\theta}(U)}} \, \xi^\theta_{(1)} + \, \xi^\theta_{(2)} \, .
\ee
In the tip limit, the linearized equations of motions for the massive bosonic modes are (see \cite{Bigazzi:2022gal})
\be
(-\eta^{\alpha\beta} \partial_\alpha \partial_\beta + M_B^2) \xi^u_{(1)} = 0 \, , \quad (-\eta^{\alpha\beta} \partial_\alpha \partial_\beta + M_B^2) \xi^\theta_{(1)} = 0 \, .
\ee
On the other hand, the second order equation for the $u$-direction requires that
\be
\eta^{\alpha\beta} \[2 \, c_\alpha \partial_\beta - \frac{2}{3\,m_0} \partial_\alpha \partial_\beta\] \xi^u_{(2)} + \frac{1}{2} \, \eta^{\alpha\beta} \( \partial_\alpha \partial_\beta - 3 \, m_0 \, c_\alpha \partial_\beta \) \[\(\xi^u_{(1)}\)^2 + \(\xi^\theta_{(1)}\)^2 \] = 0
\ee
and so fixes
\be \label{xi2u}
\xi^u_{(2)} = \frac{3}{4} \, m_0 \[\(\xi^u_{(1)}\)^2 + \(\xi^\theta_{(1)}\)^2 \] \, .
\ee
Finally, the second order equations for the Minkowskian directions are
\be \label{x2nu}
\partial_\sigma F - \, \eta^{\alpha\beta} \partial_\alpha \[ \frac{3}{m_0} \partial_\beta X^\nu \xi^u_{(2)} + \alpha' \frac{\lambda}{3} \partial_\beta \xi^\nu_{(2)} + i \, \frac{\lambda}{8} \alpha' \partial_\beta X^\nu \bar \psi \, \tilde \Gamma \, \psi \] \eta_{\mu\nu} = 0 \, .
\ee
where
\be
F = i \, \frac{\lambda}{8} \, \alpha' \[ \bar \psi \( \rho \, \Gamma_{\underline \mu} \Gamma_{\underline 1} + p^i \, \Gamma^{(11)} \Gamma_{\underline\mu \underline i} \) \tilde \Gamma \psi \] \, .
\ee
Notice that the link between the fluctuations along the $u$-direction and the ones along the $y$-directions is given by \eqref{link}. So, in the light of \eqref{cigarfluc}, in the tip limit we get
\be
\xi_{(2)}^u = \frac34 \, u_0 \, y^2 = \alpha' \, \frac\lambda8 \, m_0 \, y^2 \, .
\ee
In this way, \eqref{x2nu} reproduces \eqref{inter}.

Using these results and up to quadratic order in quantum fluctuations, a tedius computation allows us to reproduces \eqref{boseT} and so we get the same mass-shell condition as in \eqref{massshellcond}. Notice that, thanks to \eqref{xi2u}, all the dependence on the $c_\alpha$-coefficients vanishes. So, actually, there is no need to constraint them, as in \cite{Bigazzi:2022gal}
\section{Details on the fermionic sector}
\label{appF}
The equations of motion (\ref{fetot}) and the canonical Hamiltonian (\ref{hamf}) for the fermionic fields single out the operator
\be
Q =  \left[p^i \, \eta_{ij} \, \Gamma^{\underline j} + \rho \, \Gamma^{(11)}\Gamma^{\underline 1}\right]\,.
\ee
After some algebra it is possible to show that the related operator
\be
P=\frac{1}{2 \, p^0} \Gamma^{\underline 0} Q\,,
\ee
is, to quadratic order, an orthogonal projector
\be
P^2 \psi = P \psi +O(\psi^2, \xi^2)\,,\qquad P^{\dagger}=P\,.
\ee
This suggests us to redefine the fermionic fields as
\be
\tilde\psi = \frac{i}{\sqrt{2 \, p^0}} \, P \, \psi\,,
\ee
which, as it can be deduced from (\ref{fetot}), satisfy the same equations of motion (\ref{fe}) as the original $\psi$ fields. Using the above definition 
%the fermionic 
%action (\ref{ESF2}) reduces to
%\be
%S_F = \frac{i\lambda}{24\pi}\int d\tau d\sigma \tilde\psi^{\dagger}\left[\partial_{\tau}+\Gamma^{11}\partial_{\sigma} - M_F \Gamma^{11}\Gamma^0\widetilde\Gamma\right]\tilde\psi\,,
%\ee 
%so that the related 
the canonical fermionic Hamiltonian (see eq. (\ref{hamf})) gets the simple form
\be
H_F = -\frac{i\lambda}{24\pi}\int d\sigma \tilde\psi^{\dagger}\partial_{\tau}\tilde\psi\,.
\ee
This also suggests that it is convenient to work with the redefined fields when quantizing the system.

For completeness, in a similar way we obtain
\be \label{levmatch}
\int \hspace{-4pt} d\sigma \[\mathcal{T}_{\tau\sigma}^B + \mathcal{T}_{\tau\sigma}^F\] = \frac{12\pi}{\lambda} \hspace{-4pt} \int \hspace{-4pt} d\sigma \mathcal{P} \, ,
\ee
where
\be
\mathcal{P} = \frac{\lambda}{12\pi} \int d\sigma \[ \partial_\tau \xi^p_{(1)} \partial_\sigma \xi^q_{(1)} \delta_{pq} + \frac{i}{2} \tilde \psi^\dagger \partial_\sigma \tilde \psi \] \, , \quad p \, , q \neq + \, , -
\ee
is the worldsheet momentum. In other words, what we get in \eqref{levmatch} is the level-matching condition, as usual.

\section{Calculation of $\Delta(\mu)$}
\label{Delta}

In this appendix, we shall calculate the function $\Delta(\mu)$ defined in \eqref{Delta:D.def}, which appears both in the mass-shell condition \eqref{massshellcond} and in the Virasoro constraint \eqref{wbg:T.hat.vac1}. We recall that these sums are the contributions from six massless scalars, two massive scalars with mass parameter $\mu$ and eight massive 2-d spinors with mass parameter $\frac12\mu$. 

We shall use Zeta function regularization \cite{Elizalde1995} to evaluate each sum. A similar calculation involving an equal number of massive scalar and spinor fields (with equal masses) has been done in \cite{Hyun:2003ks}. Before starting, let us add a word of warning. Looking at \eqref{Delta:D.def}, one may be tempted to trade the summands with odd $n$ in the sum from the massive scalars against half of the fermion contribution. Doing so would make the contribution from the massive fields look identical to the analogue expression for four scalars and four fermions, both with mass $\frac12 \mu$. However, this is wrong because it would involve an illegal re-ordering of the infinite sums. 

Adopting Zeta function regularization, we evaluate the expression
\begin{equation}
\label{Delta:D.s}
	\Delta(s;\mu) = -6 \sum\limits_{n=1}^\infty n^{-s} - \mu^{-s} - 2 \sum\limits_{n=1}^\infty \left(n^2+\mu^2\right)^{-\frac{s}2} + 8 \sum\limits_{n=1}^\infty \left[\left(n-\frac12\right)^2+\frac{\mu^2}{4}\right]^{-\frac{s}2}~, 
\end{equation}
taking $s$ sufficiently large with the aim of performing an analytic continuation to $s= -1$. We look for a power series expansion in $\mu$. It is straightforward to obtain (see also \cite{Hyun:2003ks})
\begin{align}
\label{Delta:Ds.expand}
	\Delta(s;\mu) = - 6 \zeta(s) - \mu^{-s} -2 \sum\limits_{k=0}^\infty \frac{(-1)^k \Gamma\left(\frac{s}{2}+k\right)}{k! \Gamma\left(\frac{s}{2}\right)} \mu^{2k} \left[ \zeta(s+2k) - 2^{2-2k} \zeta(s+2k,\textstyle{\frac12}) \right]~,
\end{align}  
where we have swapped the order of the summations, which can be done under the condition that $s$ is sufficiently large. The (generalized) Riemann Zeta functions are given by 
\begin{align}
\label{Delta:zeta1}
	\zeta(s) &= \sum\limits_{n=1}^\infty n^{-s} = \zeta(s,1)~,\\
\label{Delta:zeta2}
	\zeta(s,b) &= \sum\limits_{n=0}^\infty \left(n+b\right)^{-s}~.
\end{align}
After using the useful identity 
\begin{equation}
\label{Delta:zeta.rel}
	\zeta(s,\textstyle{\frac12}) = \left(2^s-1\right) \zeta(s)~,
\end{equation}
and writing separately the terms with $k=0$ and $k=1$, \eqref{Delta:Ds.expand} becomes
\begin{align}
\label{Delta:Ds.exp2}
	\Delta(s;\mu) &= - 8(2-2^s) \zeta(s) - \mu^{-s} + 2\mu^2 s \left(1 - 2^{s+1}\right) \zeta(s+2) \\
\notag &\quad  -2 \sum\limits_{k=2}^\infty \frac{(-1)^k \Gamma\left(\frac{s}{2}+k\right)}{k! \Gamma\left(\frac{s}{2}\right)} \mu^{2k} \left[ 1 + 2^{2(1-k)} - 2^{2+s} \right]\zeta(s+2k)~.
\end{align}
Finally, after analytic continuation and using 
$$ \zeta(s) = \frac{1}{s-1} -\psi(1) +\mathcal{O}(s-1) $$
in the term with $\mu^2$, we observe that the would-be pole at $s=-1$ is compensated by the factor $(1-2^{s+1})$. Therefore, we can take the limit $s\to -1$ to obtain
\begin{equation}
\label{Delta:D.result}
	\Delta(\mu) = 1 - \mu + 2 \ln 2\, \mu^2 
-2 \sum\limits_{k=2}^\infty \frac{(-1)^k \Gamma\left(k-\frac{1}{2}\right)}{k! \Gamma\left(-\frac{1}{2}\right)} \mu^{2k} \left[2^{2(1-k)} - 1\right]\zeta(2k-1)~.
\end{equation}

Notice that the above is not divergent, despite the fact that the individual bosonic and fermionic contributions in \eqref{Delta:D.def} are divergent. In \cite{Hyun:2003ks}, the authors regularize and renormalize each of them separately. This affects the term proportional $\mu^2$ by a factor of two.

\section{The tensor $\hat{T}_{\alpha\beta}$}
\label{app:T}

In this appendix, we evaluate explicitly the tensor $\hat{T}_{\alpha\beta}$ defined in \eqref{wbg:T.hat},
\begin{equation}
\label{appT:T.hat}
	\hat{T}_{\alpha\beta} 
	= \tilde{T}_{\alpha\beta} + \left( \frac{\nabla_\alpha\nabla_\beta}{\Box} -g_{\alpha\beta} \right) \tilde{T}~.
\end{equation}
As stated elsewhere, $\tilde{T}_{\alpha\beta}$ is the canonical, conserved stress-energy tensor for the physical fluctuations, which is related to the traceless $T_{\alpha\beta}$ by \eqref{wbg:T.tilde}.  
The tensor $\hat{T}_{\alpha\beta}$ is evidently both, conserved and traceless. This implies, by standard arguments, that $\hat{T}_{+-}=0$, while $\hat{T}_{++}$ and $\hat{T}_{--}$ must be functions of $\sigma^+$ and $\sigma^-$, respectively, where $\sigma^{\pm}=\tau\pm \sigma$. Because the massive fields do not have such a dependence, the only way to meet the above statements is that the massive fields contribute a coordinate-independent (and traceless) $\hat{T}_{\alpha\beta}$. This will be confirmed by our explicit calculation. In contrast, the massless fields give rise to left- and right-mover terms that are familiar from bosonic string theory. We shall, with some abuse of notation, consider the contributions of the individual fields separately and collect all results at the end.

Let us start with the massive bosons. For a free scalar boson with mass $M$,\footnote{$M^2$ is given by \eqref{wbg:M55} and is related to $\mu^2$ by $\mu^2=|g_{\tau\tau}|M^2.$} the stress-energy tensor is 
\begin{equation}
\label{appT:TB}
	\tilde{T}_{\alpha\beta} = (\nabla_\alpha \chi)(\nabla_\beta \chi) -\frac12 g_{\alpha\beta} (\nabla_\gamma \chi)(\nabla^\gamma \chi) -\frac12 g_{\alpha\beta}M^2 \chi^2~.
\end{equation}
It is conserved on-shell, but not traceless, the trace being $\tilde{T} = -M^2 \chi^2$. 

First, using the mode expansion \eqref{wbg:scal.mode.massive} and performing normal ordering, $\tilde{T}_{\tau\tau}$ is
\begin{align}
\label{appT:Ttt}
	\tilde{T}_{\tau\tau} &= \frac12\left[ (\partial_\tau\chi)^2 + (\partial_\sigma\chi)^2 +\mu^2 \chi^2 \right]\\
\notag &= 
	\sum\limits_{m,n} \frac{\alpha'}{4\sqrt{\omega_m\omega_n}} \left\{ (\omega_m\omega_n +mn -\mu^2)
	\left[a_m a_n \e{-i[(\omega_m+\omega_n)\tau+(m+n)\sigma]}+\text{cc.}\right] \right. \\
\notag &\quad
	\left. + (\omega_m\omega_n +mn +\mu^2) \left(2a_m^\dagger a_n +\delta_{mn}\right) \e{i[(\omega_m-\omega_n)\tau+(m-n)\sigma]}\right\}~.
\end{align} 
Similarly, $\tilde{T}_{\tau\sigma}$ is
\begin{align}
\label{appT:Tts}
	\tilde{T}_{\tau\sigma} &= (\partial_\tau\chi)(\partial_\sigma\chi)\\
\notag &= 
	\sum\limits_{m,n} \frac{\alpha'n\sqrt{\omega_m}}{2\sqrt{\omega_n}} 
	\left\{a_m a_n \e{-i[(\omega_m+\omega_n)\tau+(m+n)\sigma]}+a_m^\dagger a_n \e{i[(\omega_m-\omega_n)\tau+(m-n)\sigma]}+\text{cc.}\right\}
\end{align} 
The normal ordering constants have cancelled in \eqref{appT:Tts} between positive and negative $n$. 
For the trace $\tilde{T}$, we find
\begin{align}
\label{appT:T.trace}
	\tilde{T} &= -M^2 \chi^2\\
\notag &= 
	\sum\limits_{m,n} \frac{M^2 \alpha'}{2\sqrt{\omega_m\omega_n}} 
	\left\{a_m a_n \e{-i[(\omega_m+\omega_n)\tau+(m+n)\sigma]}+\text{cc.}\right\} \\
\notag &\quad
	- \sum\limits_{m\neq n} \frac{M^2 \alpha'}{\sqrt{\omega_m\omega_n}} 
	a_m^\dagger a_n \e{i[(\omega_m-\omega_n)\tau+(m-n)\sigma]}  
	- \sum\limits_{n} \frac{M^2 \alpha'}{2\omega_n} 
			\left(2a_n^\dagger a_n +1 \right)~.
\end{align} 
Note that in the last sum we have separated the terms that are independent of $\tau$ and $\sigma$. It follows that 
\begin{align}
\label{appT:T.box}
	\frac1{\Box}\tilde{T} &= 
	\sum\limits_{m,n} \frac{\mu^2 \alpha'	\left\{a_m a_n \e{-i[(\omega_m+\omega_n)\tau+(m+n)\sigma]}+\text{cc.}\right\}}{2\sqrt{\omega_m\omega_n}[(\omega_m+\omega_n)^2-(m+n)^2]} \\
\notag &\quad
	- \sum\limits_{m\neq n} \frac{\mu^2 \alpha' a_m^\dagger a_n \e{i[(\omega_m-\omega_n)\tau+(m-n)\sigma]}}{\sqrt{\omega_m\omega_n}[(\omega_m-\omega_n)^2-(m-n)^2]} 		 
	+\frac12 \tau^2 \sum\limits_{n} \frac{\mu^2 \alpha'}{2\omega_n} 
			\left(2a_n^\dagger a_n +1 \right)~.
\end{align} 

Therefore, the last term appearing in the $\tau\tau$ component of \eqref{appT:T.hat} is
\begin{align}
\label{appT:T.box.comb}
	\left(\frac{\partial_\tau^2}{\Box}-g_{\tau\tau}\right) \tilde{T} &= 
	-\sum\limits_{m,n} \frac{\mu^2 \alpha' (m+n)^2	\left\{a_m a_n \e{-i[(\omega_m+\omega_n)\tau+(m+n)\sigma]}+\text{cc.}\right\}}{2\sqrt{\omega_m\omega_n}[(\omega_m+\omega_n)^2-(m+n)^2]} \\
\notag &\quad
	+ \sum\limits_{m\neq n} \frac{\mu^2 \alpha' (m-n)^2 a_m^\dagger a_n \e{i[(\omega_m-\omega_n)\tau+(m-n)\sigma]}}{\sqrt{\omega_m\omega_n}[(\omega_m-\omega_n)^2-(m-n)^2]}~.
\end{align} 
Adding this to \eqref{appT:Ttt} and using the identities 
\begin{align}
\label{appT:id1}
	\omega_m\omega_n +mn -\mu^2 &= -\frac12 (\omega_m-\omega_n)^2 + \frac12 (m+n)^2~,\\ 
	\omega_m\omega_n +mn +\mu^2 &= \frac12 (\omega_m+\omega_n)^2 - \frac12 (m-n)^2 
\end{align}
yields after some manipulations
\begin{equation}
\label{appT:Ttt.rhs}
	\hat{T}_{\tau\tau} = \alpha' \sum\limits_n \omega_n \left(a_n^\dagger a_n +\frac12\right)~.
\end{equation}
All oscillator off-diagonal terms have cancelled explicitly, and only a coordinate independent term remains. An analogous computation yields 
\begin{equation}
\label{appT:Tts.rhs}
	\hat{T}_{\tau\sigma} = \alpha' \sum\limits_n  n a_n^\dagger a_n~.
\end{equation}

Let us turn to the fermionic part. The canonical stress-energy tensor of a free massive fermion is\footnote{Using the Majorana condition, the last two terms in the brackets are identical to the first two terms. In this symmetric form, however, the conservation of $\tilde{T}_{\alpha\beta}$ is more obvious.}
\begin{equation}
\label{appT:T.F}
	\tilde{T}_{\alpha\beta} = \pi \alpha' i \left[ \bar{\psi} \Gamma_\alpha \nabla_\beta \psi + \bar{\psi} \Gamma_\beta \nabla_\alpha \psi -(\nabla_\beta\bar{\psi}) \Gamma_\alpha \psi -(\nabla_\alpha\bar{\psi}) \Gamma_\beta \psi\right]~.
\end{equation}  

Substituting the mode expansion \eqref{wbg:spin.mode}, one finds 
\begin{align}
\label{appT:T.tt.F}
	\tilde{T}_{\tau\tau} &= \sum\limits_{r,r'} \frac{\alpha' (\omega_r+\omega_{r'})\e{i(\omega_r-\omega_{r'})\tau}}{8\sqrt{\omega_r\omega_{r'}(\omega_r+r)(\omega_{r'}+r')}} \\
\notag &\quad 
	\times \left\{ \left[ \mu_s^2 +(\omega_r+r)(\omega_{r'}+r') \right] 
		\left[c_r^s{}^\dagger c_{r'}^s \e{i(r-r')\sigma} + d_r^s{}^\dagger d_{r'}^s \e{-i(r-r')\sigma} \right] \right. \\
\notag &\quad 
	\left. + i\mu_s \left(\omega_r +\omega_{r'} +r+r' \right)
	 	\left[c_r^s{}^\dagger d_{r'}^s \e{i(r+r')\sigma} - d_r^s{}^\dagger c_{r'}^s \e{-i(r+r')\sigma} \right]\right\}~.
\end{align}
This expression is not yet normal ordered. We postpone normal ordering to the end of the calculation.
Similarly,
\begin{align}
\label{appT:T.ts.F}
	\tilde{T}_{\tau\sigma} &= \sum\limits_{r,r'} \frac{\alpha' \e{i(\omega_r-\omega_{r'})\tau}}{8\sqrt{\omega_r\omega_{r'}(\omega_r+r)(\omega_{r'}+r')}} \\
\notag &\quad 
	\times \left\{ (r+r')(\omega_r+r)(\omega_{r'}+r')
		\left[c_r^s{}^\dagger c_{r'}^s \e{i(r-r')\sigma} - d_r^s{}^\dagger d_{r'}^s \e{-i(r-r')\sigma} \right] \right. \\
\notag &\quad 
	\left. + i\mu_s (r-r') (\omega_r+\omega_{r'}+r+r') 
	 	\left[c_r^s{}^\dagger d_{r'}^s \e{i(r+r')\sigma} + d_r^s{}^\dagger c_{r'}^s \e{-i(r+r')\sigma} \right]\right\}~.
\end{align}

The trace of \eqref{appT:T.F} is 
\begin{align}
\notag 
	\tilde{T} &= 4 \pi \alpha' i \frac{m_0}{\rho u_0^{3/4}} \mu_s\bar{\psi}^s\psi^s \\
\label{appT:T.trace.F} &=
	-\frac{m_0^2}{\rho^2 u_0^{3/2}} \sum\limits_{r,r'} \frac{\alpha' \mu_s \e{i(\omega_r-\omega_{r'})\tau}}{4\sqrt{\omega_r\omega_{r'}(\omega_r+r)(\omega_{r'}+r')}} \\
\notag &\quad 
	\times \left\{ \mu_s \left(\omega_r +\omega_{r'} +r+r' \right)  
		\left[c_r^s{}^\dagger c_{r'}^s \e{i(r-r')\sigma} + d_r^s{}^\dagger d_{r'}^s \e{-i(r-r')\sigma} \right] \right. \\
\notag &\quad 
	\left. + i\left[ \mu_s^2 +(\omega_r+r)(\omega_{r'}+r') \right] 
	 	\left[c_r^s{}^\dagger d_{r'}^s \e{i(r+r')\sigma} - d_r^s{}^\dagger c_{r'}^s \e{-i(r+r')\sigma} \right]\right\}~.
\end{align}

It is then straightforward, albeit somewhat tedious, to obtain
\begin{equation}
\label{appT:Ttt.F.rhs}
	\hat{T}_{\tau\tau} = \alpha' \sum\limits_{r>0}
	\omega_r \left( c_r^s{}^\dagger c_r^s +  d_r^s{}^\dagger d_r^s -1 \right)
\end{equation}
and
\begin{equation}
\label{appT:Tts.F.rhs}
	\hat{T}_{\tau\sigma} = \alpha' \sum\limits_{r>0} r
	\left( c_r^s{}^\dagger c_r^s -  d_r^s{}^\dagger d_r^s \right)~.
\end{equation}
The operators have been normal ordered in these expressions. As for the massive scalars, the off-diagonal elements in the oscillators have cancelled. 

It remains to look at the massless scalars. For each of these, one gets the standard CFT results
\begin{align}
\label{appT:Ttt.massless}
	\hat{T}_{\tau\tau} &= T_{\tau\tau} = \alpha' \sum\limits_n \left[ L_n \e{-in(\tau -\sigma)} + \tilde{L}_n \e{-in(\tau+\sigma)}\right]~,\\
\label{appT:Tts.massless}
	\hat{T}_{\tau\sigma} &= T_{\tau\sigma} = \alpha' \sum\limits_n \left[ -L_n \e{-in(\tau-\sigma)} + \tilde{L}_n \e{-in(\tau+\sigma)}\right]~,
\end{align}
where we have defined $\alpha_0=\tilde{\alpha}_0=q_0$, and the Virasoro generators are 
\begin{equation}
\label{appT:L}
	L_n = \frac12\sum\limits_m \alpha_m \alpha_{n-m}~,\qquad \tilde{L}_n = \frac12\sum\limits_m \tilde{\alpha}_m \tilde{\alpha}_{n-m}~.
\end{equation}
As usual, the Virasoro generators are normal orderd by
\begin{equation}
\label{appT:L.normal.order}
	L_n =\, :L_n: + \frac12 \delta_{n,0} \sum\limits_{n>0} n = \, :L_n: -\frac1{24} \delta_{n,0}~,
\end{equation}
and analogously for $\tilde{L}_n$.

Finally, let us collect all the contributions. It is obvious that also the full $\hat{T}_{\tau\tau}$ and $\hat{T}_{\tau\sigma}$ have the form \eqref{appT:Ttt.massless} and \eqref{appT:Tts.massless}, respectively. Whereas $L_n$ and $\tilde{L}_n$ for $n\neq0$ are given by summing the contributions \eqref{appT:L} for all six massless modes, $L_0$ and $\tilde{L}_0$ also receive contributions from the massive fields,
\begin{align}
\label{appT:L0.full}
	L_0 &= \frac12 q_0\cdot q_0 + \sum\limits_{n>0} \alpha_{-n}\cdot \alpha_n 
	+ \frac12 \sum\limits_n (\omega_n-n)a_n^\dagger \cdot a_n \\
\notag &\quad 
	+ \frac12 \sum\limits_{r>0}\left[ (\omega_r - r) c_r^\dagger \cdot c_r + (\omega_r + r) d_r^\dagger \cdot d_r \right] 
	 -\frac12 \Delta(\mu)~,\\
\label{appT:tildeL0.full}
	\tilde{L}_0 &= \frac12 q_0\cdot q_0 + \sum\limits_{n>0} \tilde{\alpha}_{-n}\cdot \tilde{\alpha}_n
	+ \frac12 \sum\limits_n (\omega_n+n)a_n^\dagger \cdot a_n \\ 
\notag &\quad 
	+ \frac12 \sum\limits_{r>0}\left[ (\omega_r + r) c_r^\dagger \cdot c_r + (\omega_r - r) d_r^\dagger \cdot d_r \right] 
	 -\frac12 \Delta(\mu)~.
\end{align}
Here, $\Delta(\mu)$ collects all the normal ordering constants and is given by \eqref{Delta:D.def}, while the product dots indicate  summations over all fields.

\end{appendix}

\bibliographystyle{JHEP}
\bibliography{hagedorn}

\end{document}